\documentclass[sigconf,nonacm]{acmart}

\usepackage{tabularx}
\usepackage{booktabs}
\usepackage{graphicx}
\usepackage{subcaption}
\usepackage{balance}
\usepackage{multirow}
\usepackage{pifont} % For checkmarks and crosses
\usepackage{colortbl} % For coloring cells
\usepackage{float}
\usepackage{tikz}
\usepackage[normalem]{ulem}
\usepackage{array}
\usepackage[table,xcdraw,dvipsnames]{xcolor}  % for colored marks
\usepackage{makecell}  % For multi-line cell content
\usepackage{amsmath}    % for \text and align environments
\usepackage{amsfonts}   % for \mathbb
\usepackage{hyperref}
\usepackage{enumitem}
\usepackage{xurl}
\usepackage{placeins}
\usepackage{afterpage}

\usepackage{array}
\newcolumntype{L}[1]{>{\raggedright\arraybackslash}p{#1}}

\usepackage[normalem]{ulem}

\usepackage{algorithm}
\usepackage{algpseudocode}

\AtBeginDocument{%
  }

\setcopyright{none}
\renewcommand\footnotetextcopyrightpermission[1]{}
\newcommand{\cmark}{\textcolor{ForestGreen}{\ding{51}}}
\newcommand{\xmark}{\textcolor{BrickRed}{\ding{55}}}

\newcommand{\scbf}[1]{\vspace {4pt}\noindent{\textbf{#1.}}}
\newcommand\RansomSaver{\emph{Rhea}}

\begin{document}

\title{\RansomSaver{}: Detecting Privilege-Escalated Evasive Ransomware Attacks Using Format-Aware Validation in the Cloud}

\author{Beom Heyn Kim}
%\authornote{Corresponding Author. Email: beomheynkim@hanyang.ac.kr}
\authornote{Corresponding Author}
%\authornote{Authors are affiliated with the Department of Computer Science \& Engineering (Major in Bio Artificial Intelligence).}
\email{beomheynkim@hanyang.ac.kr}
\orcid{0000-0002-8650-6082}
\affiliation{%
  \institution{Hanyang University ERICA}
  \city{Ansan}
  \state{Gyeonggi}
  \country{Republic of Korea}
}

\author{Seok Min Hong}
\email{hsmint@hanyang.ac.kr}
\orcid{1234-5678-9022}
\affiliation{%
  \institution{Hanyang University ERICA}
  \city{Ansan}
  \state{Gyeonggi}
  \country{Republic of Korea}
}

\author{Mohammad Mannan}
\email{m.mannan@concordia.ca}
\orcid{1234-5678-9012}
\affiliation{%
  \institution{Concordia University}
  \city{Montreal}
  \state{Quebec}
  \country{Canada}
}

%\renewcommand{\shortauthors}{Anonymized}
\renewcommand{\shortauthors}{Kim et al.}

\begin{abstract}
  % A clear and well-documented \LaTeX\ document is presented as an
  % article formatted for publication by ACM in a conference proceedings
  % or journal publication. Based on the ``acmart'' document class, this
  % article presents and explains many of the common variations, as well
  % as many of the formatting elements an author may use in the
  % preparation of the documentation of their work.
  Ransomware variants increasingly combine privilege escalation with sophisticated evasion strategies such as intermittent encryption, low-entropy encryption, and imitation attacks. Such powerful ransomware variants, \emph{privilege-escalated evasive ransomware} (PEER), can defeat existing solutions relying on I/O-pattern analysis by tampering with or obfuscating I/O traces. Meanwhile, conventional statistical content-based detection becomes unreliable as the encryption size decreases due to sampling noises. We present \RansomSaver{}, a cloud-offloaded ransomware defense system that analyzes replicated data snapshots, so-called \emph{mutation snapshots}. \RansomSaver{} introduces \emph{Format-Aware Validation} that validates the syntactic and semantic correctness of file formats, instead of relying on statistical or entropy-based indicators. By leveraging file-format specifications as detection invariants, \RansomSaver{} can reliably identify fine-grained and evasive encryption even under elevated attacker privileges. Our evaluation demonstrates that \RansomSaver{} significantly outperforms existing approaches, establishing its practical effectiveness against modern ransomware threats. 
\end{abstract}



\maketitle

\begin{center}
{\small \textit{Preprint — under review (January 2026)}}
\end{center}

\section{Introduction}\label{sec:intro}

Ransomware increasingly cause severe operational and financial damage across various domains. In 2024, Ascension, which is one of the largest U.S.\ healthcare providers, suffered from a ransomware attack leading to the disruption of clinical operations and compromised records of 5.6 million patients and staff~\cite{ransomware-attack-2024}. The rise of ransomware-as-a-service (RaaS) groups such as Medusa has further amplified this trend~\cite{ransomware-medusa-attack-2025}. By 2025, Medusa has been used to compromise more than 300 victims, encompassing healthcare, education, insurance, and other critical domains. 
What makes the situation worse is that the modern ransomware campaigns increasingly rely on privileged execution techniques, such as Bring-Your-Own-Vulnerable-Driver (BYOVD) attacks~\cite{techradar_killer_tool_2025,techradar_akira_defender_2025}. Once adversaries gain root privileges, it is not too difficult for attackers to disable security monitors. In parallel, advanced ransomware attacks often employ sophisticated evasion strategies including intermittent encryption~\cite{ransomware-intermittent-encryption-2023,erw-radar}, low-entropy encryption, and imitation attacks~\cite{animagus}. These trends collectively produce a class of \emph{Privilege-Escalated Evasive Ransomware (PEER)}.

Unfortunately, existing ransomware defenses offer limited protection against PEER adversaries. Previous proposals tend to heavily rely on I/O-pattern analysis at the user-, kernel-, hypervisor-, or hardware-level~\cite{unveil,cryptodrop,redemption,shieldfs,peeler,ransomtag,ransomspector,ssd-insider,flashguard,erw-radar,ransomblocker,AMOEBA}. However, I/O-pattern analysis becomes largely untrustworthy under PEER attacks because monitors can be disabled by privileged attackers or neutralized by obfuscated low-level traces. 

Some systems incorporate content-based checks~\cite{flashguard,erw-radar,ransomblocker,AMOEBA}. Yet, it is challenging to apply these mechanisms as the sole first-line defense on end-user devices. 
Content-based detectors partitioned files into chunks and independently analyzed each chunk using statistical methods such as entropy, $\chi^2$ statistics, or machine learning. However, these detectors inherit a fundamental statistical limitation: as the inspected window becomes smaller, the empirical byte distribution is overwhelmed by sampling noise. With the sampling noise, the observable statistical features of plaintext, compressed data, and ciphertext rapidly converge and become very difficult to reliably distinguish. Exploiting this inherent limitation of statistical approaches, attackers can evade content-based detection by encrypting only very small spans scattered throughout the file~\cite{SRLabs2024BlackBastaBuster}, which we call \emph{fine-grained partial encryption}. 
%When each encrypted span occupies only a small portion of a window, its statistical footprint is masked by surrounding plaintext, allowing the encrypted span to remain undetected. Thus, fine-grained partial encryption exploits a universal blind spot of window-based content detectors, bypassing them until substantial damage accumulates.

To overcome the limitation of previous content-based detection, we introduce \emph{format-aware validation} (FAV), which analyzes not raw byte statistics but the syntactic and semantic invariants of the underlying file format. FAV builds upon a foundational observation that \emph{a file format can serve as the correctness specification of the valid data state transition}. More specifically, the benign mutation of high-value user files such as text, zip, ooxml, pdf files should always conform to the rich syntactic and semantic structure of the file format. Even a small encrypted span typically violates those invariants, producing malformed text, broken markup, or nonsensical content. Thus, FAV can reliably identify fine-grained partial encryption that is invisible to prior content-based approaches relying on statistical methods. 
%For immutable or rarely modified binary assets such as images or media files, which may lack such structure, \RansomSaver{} applies trusted hash whitelisting to validate integrity without depending on syntactic cues. 

%\scbf{Cloud-Offloaded Defense}
In this work, we develop \RansomSaver{}, a ransomware defense architecture that offloads both detection and recovery from end devices to the cloud. Cloud offloading offers several advantages: (1) isolation from compromised hosts, including privilege-escalated attacks; (2) abundant storage for long-term retention of data snapshots; (3) ample compute capacity for fine-grained content inspection; and (4) reduced performance and management burden on end devices. \RansomSaver{} confines PEERs within virtual machines, using VMs as sandboxes to prevent ransomware from escaping the guest environment. \RansomSaver{} periodically performs a procedure called a \emph{scheduled checkpoint}, during which it replicates block-level data snapshots---referred to as \emph{mutation snapshots}---from these VMs to the cloud at designated times (e.g., nightly validation windows or expected idle periods). \RansomSaver{} then analyzes exclusively the content captured in mutation snapshots using format-aware validation (FAV), rather than relying on ephemeral and potentially untrustworthy I/O streams or coarse statistical content-based heuristics.

\scbf{Implementation and Results}
We evaluate a prototype implementation of \RansomSaver{} using a combination of
controlled partial-encryption simulations and format-diverse real-world datasets.
Across a wide range of fine-grained encryption patterns—including skip-step and
micro-granularity attacks down to single-byte modifications—\RansomSaver{} achieves
nearly 100\% detection accuracy for structured file formats such as TXT, PDF, OOXML, and ZIP.
In these formats, format-aware validation (FAV) consistently detects encryption-induced
violations even when statistical signals are intentionally minimized.
Overall, \RansomSaver{} substantially outperforms entropy- and $\chi^2$-based detectors
under fine-grained partial encryption, demonstrating that specification-level correctness
is a stronger and more reliable signal than traditional statistical methods.

\scbf{Contributions}
To the best of our knowledge, \RansomSaver{} is the first ransomware detection system that treats file format specifications as security invariants and detects encryption as a violation of those invariants rather than a statistical anomaly. Specifically, this paper makes the following contributions:
\begin{enumerate}
    \item We identify fundamental limitations of existing ransomware defenses under the PEER threat model, particularly in the presence of fine-grained partial encryption.
    \item We present a cloud-offloaded ransomware defense architecture that enables rich content analysis without imposing on-device overhead.
    \item We design and implement a novel FAV-based detection pipeline that is resilient to fine-grained partial encryption.
    \item We empirically evaluate \RansomSaver{} against both simulated and real-world ransomware, demonstrating accurate and practical detection.
\end{enumerate}

We discuss related work in Section~\ref{sec:related}, describe our threat model in Section~\ref{sec:model}, present our system design in Section~\ref{sec:design}, summarize implementation details in Section~\ref{sec:implement}, and evaluate our prototype in Section~\ref{sec:eval}. Section~\ref{sec:discuss} outlines limitations and future work, and Section~\ref{sec:conclude} concludes.

\section{Related Work}\label{sec:related}

\begin{table*}[t]
\centering
\small
\renewcommand{\arraystretch}{1.3}
\caption{Comparison of related approaches to ransomware countermeasures. Each column shows whether the corresponding feature is supported. A \cmark{} indicates support for the feature, while \xmark{} denotes the lack of such support.}
\begin{tabular}{>{\bfseries}l L{5cm} c c c c c }
\toprule
\textbf{Level} & \textbf{Approach} & 
\makecell{Tamper-\\Resistance} & \makecell{Interface\\Reduction} & 
\makecell{Software-only\\Solution} & 
\makecell{Cloud-based\\Detection} &
\makecell{Format-aware\\Validation}
 \\
\midrule

\multirow{2}{*}{OS} 
& Redemption~\cite{redemption}, ShieldFS~\cite{shieldfs}, UNVEIL~\cite{unveil}, Peeler~\cite{peeler}, CryptoDrop~\cite{cryptodrop}, RWGuard~\cite{rwguard}, PayBreak~\cite{paybreak}, CANCAL~\cite{cancal}, MS-IDS~\cite{ms-ids}, ERW-Radar~\cite{erw-radar}
& \xmark & \xmark & \cmark & \xmark & \xmark \\

\midrule

\multirow{3}{*}{Hardware} 
& SSD-Insider~\cite{ssd-insider}, FlashGuard~\cite{flashguard}, AMOEBA~\cite{AMOEBA}, RansomBlocker~\cite{ransomblocker}, LAST~\cite{LAST}
& \cmark & \xmark & \xmark & \xmark & \xmark \\
\cmidrule(lr){2-7}
& Inuksuk~\cite{inuksuk}, RSSD~\cite{rssd}
& \cmark & \cmark & \xmark & \xmark & \xmark \\

\midrule

\multirow{3}{*}{Hypervisor} 
& RansomSpector~\cite{ransomspector}, RansomTag~\cite{ransomtag}, Rocky~\cite{rocky}, 
Time Machine~\cite{time_machine}
& \cmark & \cmark & \cmark & \xmark & \xmark \\
\cmidrule(lr){2-7}
& \textbf{Rhea (This Work)} 
& \cmark & \cmark & \cmark & \cmark & \cmark \\

\bottomrule
\end{tabular}
\label{tab:ransomware_defense_comparison}
\end{table*}

Table~\ref{tab:ransomware_defense_comparison} summarizes and compares various previous proposals that have explored various countermeasures for ransomware attacks. 
Note that none of existing solutions is resilient to fine-grained partial encryption which \RansomSaver{} can detect through FAV. The rest of this section describes the related works in more details. 

\scbf{OS-level Solutions} 
OS-level solutions for ransomware defense include both user-level and kernel-level solutions. Existing detection methods focus on identifying ransomware activity through I/O pattern analysis or content-based detection. For example, UNVEIL~\cite{unveil}, RWGuard~\cite{rwguard} and Peeler~\cite{peeler} analyze I/O patterns to detect ransomware. CryptoDrop~\cite{cryptodrop}, on the other hand, uses content-based detection methods to protect user data from cryptographic ransomware. Meanwhile, Redemption~\cite{redemption}, ShieldFS~\cite{shieldfs}, and PayBreak~\cite{paybreak} provide a solution for detection, recovery, and prevention.

Recently, several detection-centric OS-level solutions have been proposed. 
ERW-Radar~\cite{erw-radar} shows that imitation attacks utilizing few predefined templates result in repeated I/O patterns. Then, it presents a detection method for imitation attacks that monitors and analyzes I/O events at the kernel-level.
CANCAL~\cite{cancal} and MS-IDS~\cite{ms-ids} explore the optimization for the detection overhead imposed on the host system. CANCAL introduces a streamlined deception-based detection approach for industrial environments where real-time detection is crucial. To attract adversaries, it strategically deploys decoy files at the path containing information-rich user data. Only when the modification of decoy files are identified, heavier in-depth ransomware screening process begins. 
Meanwhile, MS-IDS prioritizes user-level processes based on their suspicion levels and it applies the most cost-effective monitoring strategy for each. As a result, benign processes are only lightly monitored, reducing performance overhead by approximately a factor of ten.

\textit{Differences with OS-level Solutions:} A major limitation of all OS-level solutions is that those are vulnerable to privilege-escalated attacks, because attackers with elevated privileges can easily bypass or disable detection and defense mechanisms in the user-level or kernel-level~\cite{ransomhub2024,mallox2024,cl0p2025,ghost2025}
In contrast, \RansomSaver{} offers better resilience to privilege-escalated attacks, since its defense mechanism contains privileged attackers in VMs. 

\scbf{Hardware-level Solutions}
SSD-centric solutions integrate ransomware detection mechanisms directly within the SSDs at the firmware level. These solutions are more resilient to privilege-escalated attacks than OS-level solutions. AMOEBA~\cite{AMOEBA} detects ransomware attacks via content-based analysis. In addition, SSD-Insider~\cite{ssd-insider} offers mechanisms for not only detection but also recovery within SSDs. RansomBlocker~\cite{ransomblocker} proposes a low-overhead solution to protect against ransomware and LAST~\cite{LAST} proposes enabling retention-aware versioning to prevent premature deletion. Meanwhile, RSSD~\cite{rssd} focuses on hardware-isolated network-storage for post-attack analysis. FlashGuard~\cite{flashguard} leverages SSDs' intrinsic properties to defend against privileged ransomware attacks. 

There exist hardware-level solutions that are not SSD-centric. A notable example is Inuksuk~\cite{inuksuk} that employs Trusted Execution Environments (TEEs) to protect user data against tampering attacks. By leveraging hardware root-of-trust, such solutions can guarantee data integrity even against privilege-escalated attacks.

\textit{Differences with Hardware-level Solutions:} Hardware-level solutions rely on special hardware posing barriers to deployment across diverse machine configurations. SSD-centric solutions are limited to devices equipped with specialized SSDs and customized firmware, and TEE-based solutions require the presence of trusted computing hardware. 
In addition, SSD interfaces such as \textit{trim} commands~\cite{rssd} may be exposed to privilege-escalated ransomware, potentially allowing attackers to bypass built-in security measures and delete data before starting encryption. 
Also, TEE-based solutions only provides immutable backups but not an explicit detection method~\cite{inuksuk,rssd}. 
\RansomSaver{} is not restricted such hardware requirement, does not expose such interface to guest VMs, introduces a new detection method complement to immutable backup infrastructure. 
%Also, it is challenging for SSD-centric solutions to detect low-entropy attacks, where encrypted content is sparsely distributed within a file to maintain low entropy at the page level~\cite{AMOEBA,ransomblocker}. 
%TEE-based solutions only provides immutable backups but not an explicit detection method~\cite{inuksuk,rssd}. In contrast, \RansomSaver{} introduces a new detection mechanism built atop immutable backup infrastructure. 
%These hardware-based solutions can therefore serve as a complement to \RansomSaver{}.  

\scbf{Hypervisor-level Solutions}
RansomSpector~\cite{ransomspector} and RansomTag~\cite{ransomtag} employ virtual machine introspection (VMI) along with I/O-pattern analysis to gain the internal knowledge of virtual machines for more accurate ransomware detection and recovery. Other hypervisor-level solutions focus only on providing backup infrastructure~\cite{rocky,time_machine} by interposing virtual I/O and replicating backups to the Cloud.

\textit{Differences from Hypervisor-Level Solutions:} 
RansomTag~\cite{ransomtag} and RansomSpector~\cite{ransomspector} rely on I/O-pattern analysis for ransomware detection, making them inherently vulnerable to imitation attacks. 
%In addition, these solutions suffer from fundamental limitations in detecting low-entropy attacks---statistical methods for measuring the entropy of write buffers become increasingly unreliable as the buffer size decreases~\cite{encod}. 
Meanwhile, Rocky~\cite{rocky} and Time Machine~\cite{time_machine} do not incorporate effective detection mechanisms offloaded to the cloud. 
%As a result, existing hypervisor-level approaches fall short in defending against PEER attacks, which deliberately evade detection by spreading encrypted payloads across time and across locations. 
Unlike previous hypervisor-level solutions, \RansomSaver{} introduces a novel format-aware validation without relying on I/O behaviors.

%\textit{Differences with Hypervisor-level Solutions:} Introspection-centric approaches such as RansomTag~\cite{ransomtag} and RansomSpector~\cite{ransomspector} rely on I/O-based detection, making them inherently vulnerable to imitation attacks. Also, those solutions face significant limitations in detecting low-entropy attacks---statistical methods to measure the entropy of write buffers fundamentally become unreliable as the buffer size decreases~\cite{encod}.  Meanwhile, storage-centric systems such as Rocky~\cite{rocky} and Time Machine~\cite{time_machine} do not incorporate effective detection mechanisms offloaded to the cloud. As a result, existing hypervisor-level solutions fall short in defending against PEER, which deliberately evades detection by spreading its encrypted payload across time and locations. Unlike previous hypervisor-level approaches, \RansomSaver{} introduces a novel cloud-offloaded detection mechanism capable of tracking cumulative content changes and reliably detecting aforementioned evasive ransomware attacks using dual-level content-only analysis without relying on I/O behaviors.

\scbf{Miscellaneous} 
Machine learning-based anomaly detection has also been explored to detect malicious behaviors, including ransomware. Techniques such as deep learning on system logs~\cite{deeplog}, autoencoder-based network intrusion detection~\cite{kitsune}, and dataflow anomaly detection~\cite{dataflow-anomaly-detection} have been applied to identify irregular system behaviors. 
Some studies have proposed using machine learning to classify encrypted traffic and differentiate it from other high-entropy data formats. For instance, statistical and machine learning techniques has been applied to identify encrypted data~\cite{encod, cha2016detecting}. 

\textit{Differences with Miscellaneous Solutions:} Machine learning-based anomaly detection approaches are ineffective against PEER that can mimic normal I/O patterns or operate intermittently to avoid detection. Machine learning-based classifiers differentiating ciphertext from plaintext alone do not provide a comprehensive ransomware defense. 

\section{Threat Model and Assumptions}
\label{sec:model}

\RansomSaver{} assumes an enterprise VDI setting in which VMs participate in \emph{scheduled checkpoints}. Checkpoints are typically executed during periods of low user activity—such as nightly maintenance windows or idle intervals when there is no user interaction—to minimize disruption. At each scheduled checkpoint, applications are expected to finalize ongoing updates to protected user data, including any file-format–specific metadata finalization, and then enter a quiesced state in which no further writes are issued. The VM subsequently enforces a flush barrier (e.g., \texttt{sync} and, if necessary, a brief filesystem or volume quiescence) before \RansomSaver{} takes a snapshot. This process ensures that each checkpoint snapshot captures only committed, format-consistent file versions rather than transient states in the middle of unfinished updates, which are a major source of false positives for structured formats such as ZIP and OOXML. Between consecutive checkpoints, \RansomSaver{} may additionally record block-level I/O activity to support post-detection analysis and recovery, such as localizing the affected regions and reconstructing the last known-good state.

Our target system involves a user device running an operating system environment in a virtual machine (VM). The enterprise environment where VDI is frequently used is the primary target environment~\cite{vdi-adoption-trends-2025}, but it is not restricted. Any computing device capable of running a lightweight virtualization layer and connecting to cloud servers over the network can be a target environment. As mentioned earlier, such devices encompass PCs~\cite{VirtualBox}, mobile devices~\cite{AOSP_AVF_Overview}, and edge servers~\cite{cloudlet2009}. 

Our solution is not tightly coupled with cloud deployment models but requires users to trust the cloud servers, assuming adversaries cannot compromise those. If the cloud servers are protected with trusted computing hardware and users can trust the hardware providers, then a public cloud can be used~\cite{safekeeper,cks}. Otherwise, the enterprises may operate cloud servers on the on-premise private cloud. Even hybrid cloud can be used by keeping the backup of encrypted block snapshots on the public cloud while analyzing block snapshots on the private cloud. 

On the other hand, adversaries can compromise a user's operating system environment by tricking users into clicking on suspicious links in emails or visiting phishing sites, which can lead to the installation of powerful malware within the user's VM. Rootkits are especially dangerous as they can exploit vulnerabilities in the OS kernels to elevate privileges. Such privileged malware can defeat host-based antivirus software or agents for cloud antivirus services. 
However, we assume that adversaries cannot break out of the VM, as the lightweight virtualization layer has a much smaller trusted computing base (TCB). It is worth noting that not every user device is equipped with specially designed SSDs or trusted computing hardware, while a hypervisor can run on a much wider spectrum of user devices, even including smartphones~\cite{VectrasVM2024,UTM2024}.

%Malware used by adversaries may be polymorphic, with signatures not known in advance. Thus, malware installed on user devices may evade host-based antivirus software. Rootkits are especially dangerous as they can exploit vulnerabilities in the OS kernels or mount successful privilege escalation attacks. Such privileged malware can defeat host-based antivirus software or agents for cloud antivirus services. Eventually, ransomware may be installed, encrypting the user's data. Installed ransomware may be not only  traditional ransomware variants but also PEER variants.

We target ransomware leveraging standard cryptographic algorithms to encrypt user data, resulting in high-entropy ciphertext. 
That is, we exclude ransomware variants that employ custom or nonstandard encryption algorithms, which may produce low-entropy ciphertext~\cite{Bang2024Entropy,McIntosh2019Inadequacy,Han2020Effectiveness}. Such schemes are often cryptographically weak and have frequently been broken or fully decrypted through analytic efforts (e.g., NoMoreRansom~\cite{nomoreransom}), in contrast to AES-grade encryption used by modern, high-impact ransomware families.
% Also, ransomware may produce small encrypted spans to obfuscate its overall entropy during measurement attempts. For instance, it may encrypt only a few blocks within a page of a file or a write buffer.  
Additionally, we do not consider ransomware that merely restricts access to system resources without employing encryption or deletion, such as screen lockers~\cite{young2025recovering,farooq_screenlocker}. 
Also, ransomware variants stealing data and threatening disclosure(doxware/leakware~\cite{ransomware-types-by-xcitium-2025}) are out of scope.
%Also, we suppose that PEER variants keep encrypted data locally on compromised devices

%Also, we suppose that PEER variants keep encrypted data locally on compromised devices to minimize detection and optimize efficiency instead of transmitting over the network. This approach reduces network bandwidth consumption, helping PEER evades detection by security solutions that monitor outbound traffic. By encrypting data in place and avoiding excessive communication with external servers, PEER can operate covertly, prolonging its presence on infected systems and increasing its chances of successfully extorting victims.

Note that the proposed approach does not work without encryption keys if the disk is encrypted using full-disk encryption software (e.g., Windows BitLocker~\cite{bitlocker2025}). 
If blocks are encrypted by such software, \RansomSaver{} cannot distinguish whether encrypted blocks are the result of ransomware or legitimate disk encryption. 
Users have two options: (1) disabling disk encryption inside the guest while letting the underlying hypervisor encrypt their data at rest and (2) providing recovery keys to \RansomSaver{} via cloud services such as Azure Disk Encryption~\cite{Microsoft_AzureDiskEncryptionOverview2025} or on-premise Configuration Manager~\cite{Microsoft_DeployBitLockerManagementAgent2023}. 
%Alternatively, for PCs, laptops, and smartphones, disk encryption can be enabled at the hypervisor level instead.

\section{\RansomSaver{} Design}\label{sec:design}

\subsection{Challenges in Detecting PEER}\label{subsec:design_attack}

%Relying on the analysis of file-system I/O patterns to detect ransomware is highly vulnerable to evasion. 
I/O-pattern analysis becomes highly unreliable under PEER attacks.
As PEER can bypass kernel-level defenses, any file I/O traces gathered in the kernel are not trustworthy. 
This confines the trusted telemetry to block-level I/O captured at the virtualization or hardware level. 
However, detecting PEER at this layer suffers from several challenges. 
With intermittent encryption, the relevant block I/O spans grow over time as encrypted writes are dispersed across the timeline. 
Analyzing such long traces is a daunting task, because the number of block I/O events to store and process can be massive. 
Moreover, PEER can camouflage malicious block I/O by arbitrarily interleaving it with benign I/O. 
PEER can further increase its stealthiness through imitation attacks, which exponentially complicates correlating attack-related I/O events.

% Relying on I/O Behavior to detect ransomware attacks is very sensitive to evasive techniques. PER can bypass any kind of kernel-level detection, therefore any file I/O traces obtained by a kernel-level component is not trustwhorthy. This limits the trusted source of information being block I/O traces obtained at the virtualization or hardware level. However, there are a couple of challenges involved in detecting PER attacks with block-level I/O traces. For intermittent encryption, block I/O traces to analyze grows in size as the encrypted writes are spread across the timeline. Yet, analyzing long block I/O traces is a daunting task, because there can be massive number of block I/O events to store and process. Additionally, PER can masquerade attacking block I/O by mixing those with benign block I/O in a virtually unlimited way. PER can further enhance their stealthiness with a technique such as imitating encryption, thereby exponentially increasing the difficulties to correlate attacking I/O events. 

Content-based detection may sound like a promising alternative to file-I/O analysis, but it is unreliable for several challenges. 
First, fine-grained partial encryption attacks are increasingly prevalent. For example, contemporary evasive ransomware variants such as \textit{Black Basta}, \textit{BlackCat}, \textit{Play}, and \textit{LockBit} need not encrypt entire files; they flexibly adjust the offsets and sizes of encryption tasks. 
For example, \textit{Black Basta} is known to (1) encrypt the whole file for files smaller than 5000\,B; (2) iteratively encrypt 64\,B and skip 128\,B for files larger than 5000\,B but smaller than 1\,GB; and (3) encrypt first 5000\,B followed by iteratively encrypting 64\,B and skipping 6336\,B for files larger than at least 1\,GB~\cite{SRLabs2024BlackBastaBuster}.
%In addition, the research prototype ransomware \textit{ANIMAGUS} encrypts less than a fixed percentage (e.g., 50\%) of randomly chosen blocks in a file while keeping write-buffer entropy low~\cite{animagus}. 
Second, evasive variants can intentionally insert low-randomness padding alongside encrypted data. As a result, many write buffers contain mixtures of ciphertext and plaintext, causing their measured entropy to remain low enough to avoid detection. 
%Third, numerous benign files produce high-entropy content due to compression, yielding false alarms. 
Third, more fundamentally, statistical methods commonly used to measure the entropy, such as Shannon entropy or $\chi^2$ tests, are unreliable in correctly classifying plaintext and ciphertext data at small granularities (e.g., 64\,B). 
In short, when evasive attacks encrypt only small chunks at a time, distinguishing ciphertext from compressed plaintext using byte-level statistics alone is extremely difficult. To overcome these challenges, \RansomSaver{} introduces a novel technique, FAV, in its content-only detection pipeline.

\subsection{Architecture Overview}\label{subsec:design_arch}

\begin{figure}
    \includegraphics[trim=0cm 3cm 0cm 0cm,width=\linewidth]{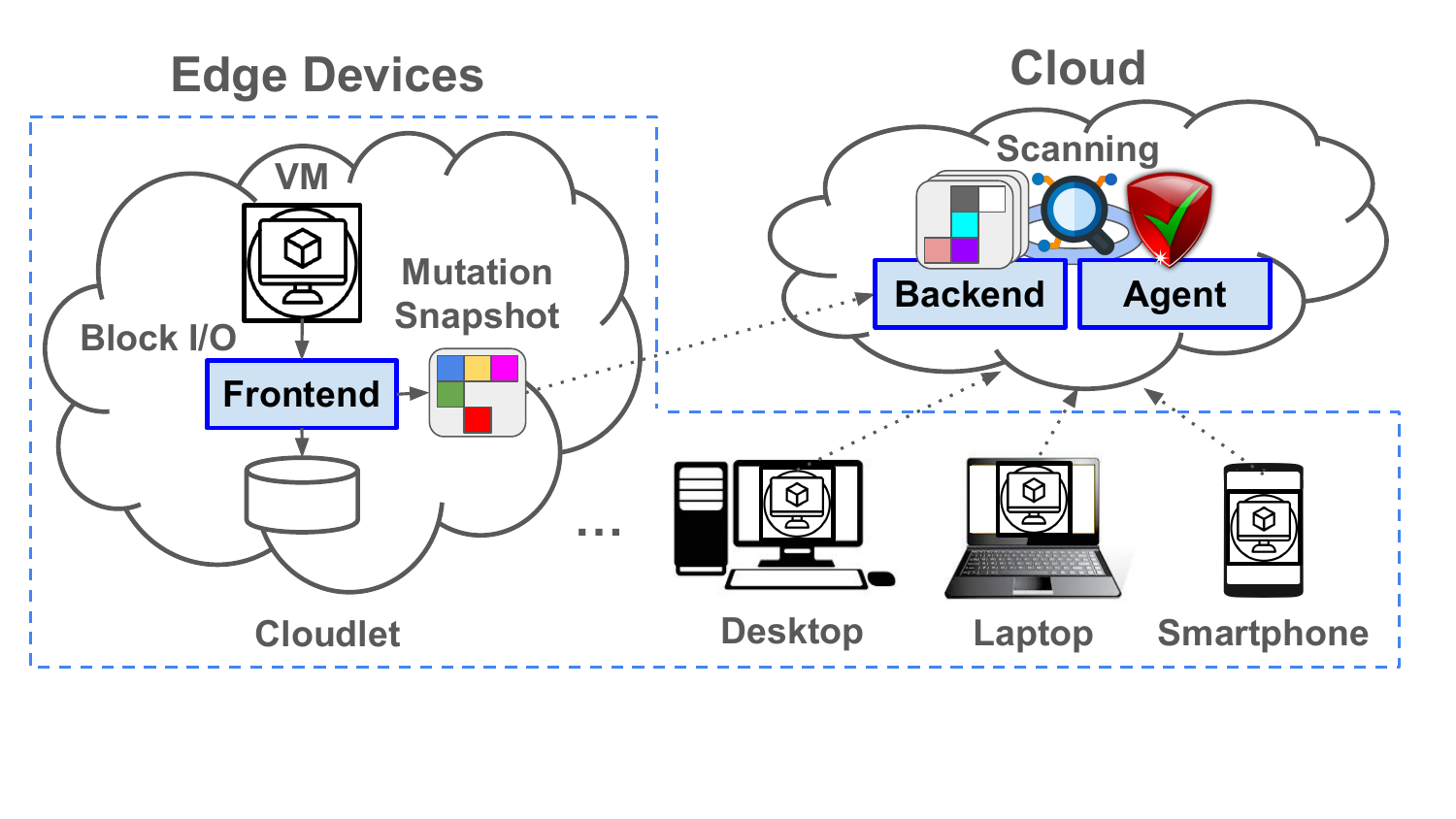}
    \caption{\RansomSaver{} architecture. 
    \RansomSaver{} \emph{frontend} periodically generates a mutation snapshot. 
    Then, \RansomSaver{} \emph{backend} stores the series of mutation snapshots as long-term immutable backups. 
    With this, \RansomSaver{} \emph{agent} conducts deeper analysis to detect ransomware attacks.
    %Agents run on edge devices such as cloudlets, laptops, smartphones, desktops, and so on.
    %\Description{RansomSaver agent creates block snapshots
    %and periodically replicate to cloud servers where 
    %cloud antivirus conducts antivirus scanning.
    }
    \vspace{-0.3cm}
    \label{fig:ransomsaver_arch}
    %\vspace{0.2cm}
\end{figure}

Figure~\ref{fig:ransomsaver_arch} illustrates the architecture overview of \RansomSaver{}. A \RansomSaver{} \emph{frontend} operates on edge devices, and a \RansomSaver{} \emph{backend} runs on the cloud as a backend server that manages a shared data storage for all connected edge devices. Edge devices can range from personal computing devices such as desktops, laptops, tablets, and smartphones to servers deployed at the cloudlets, which are basically the edge data centers sitting between user's devices and the cloud servers. The \RansomSaver{} \emph{frontend} includes a virtual block device, enabling seamless storage integration for virtual machines (VMs). When a VM runs a guest operating system, all file input/output (I/O) operations initiated within the guest OS are translated into block I/O operations, which are then processed by the underlying passthrough block device exposed by the \RansomSaver{} \emph{frontend}. 

To maintain data consistency and reliability, the \RansomSaver{} \emph{frontend} replicates data state transitions from the edge device to the \RansomSaver{} backend at \emph{scheduled checkpoints}. At the block layer, a block is the basic unit of I/O, and write requests issued by the VM are applied to the underlying block device on the edge device. During each checkpoint interval (epoch), \RansomSaver{} tracks block-level modifications and retains the most recent version of each modified block as a \emph{block snapshot} for that epoch. At the end of the epoch—after the system enforces a flush and quiescence barrier to ensure that all pending updates are finalized—these block snapshots are aggregated into a \emph{mutation snapshot}, which captures the complete data state transition that occurred during the epoch. The resulting mutation snapshot is then replicated to the \RansomSaver{} \emph{backend}, where it is stored on a distributed storage system such as DynamoDB~\cite{dynamodb}. Because a mutation snapshot represents the committed block state at the checkpoint boundary, replaying mutation snapshots in sequence allows the backend to faithfully reconstruct block-device snapshots in the cloud.

% The \RansomSaver{} \emph{backend} stores mutation snapshots on distributed storage systems such as DynamoDB~\cite{dynamodb}, ensuring accessibility from multiple \RansomSaver{} \emph{frontend}s. 
% Any mutation snapshot recorded in the past can be retrieved by another \RansomSaver{} \emph{frontend}. 
% Furthermore, all data written by any process within a VM over the epochs in the past will be preserved through the immutable series of mutation snapshots, enabling reliable data recovery and data loss prevention not only against failures but also against tampering attacks of ransomware.

\begin{figure}
    \includegraphics[trim=2cm 1cm 2cm 0cm,width=\columnwidth]{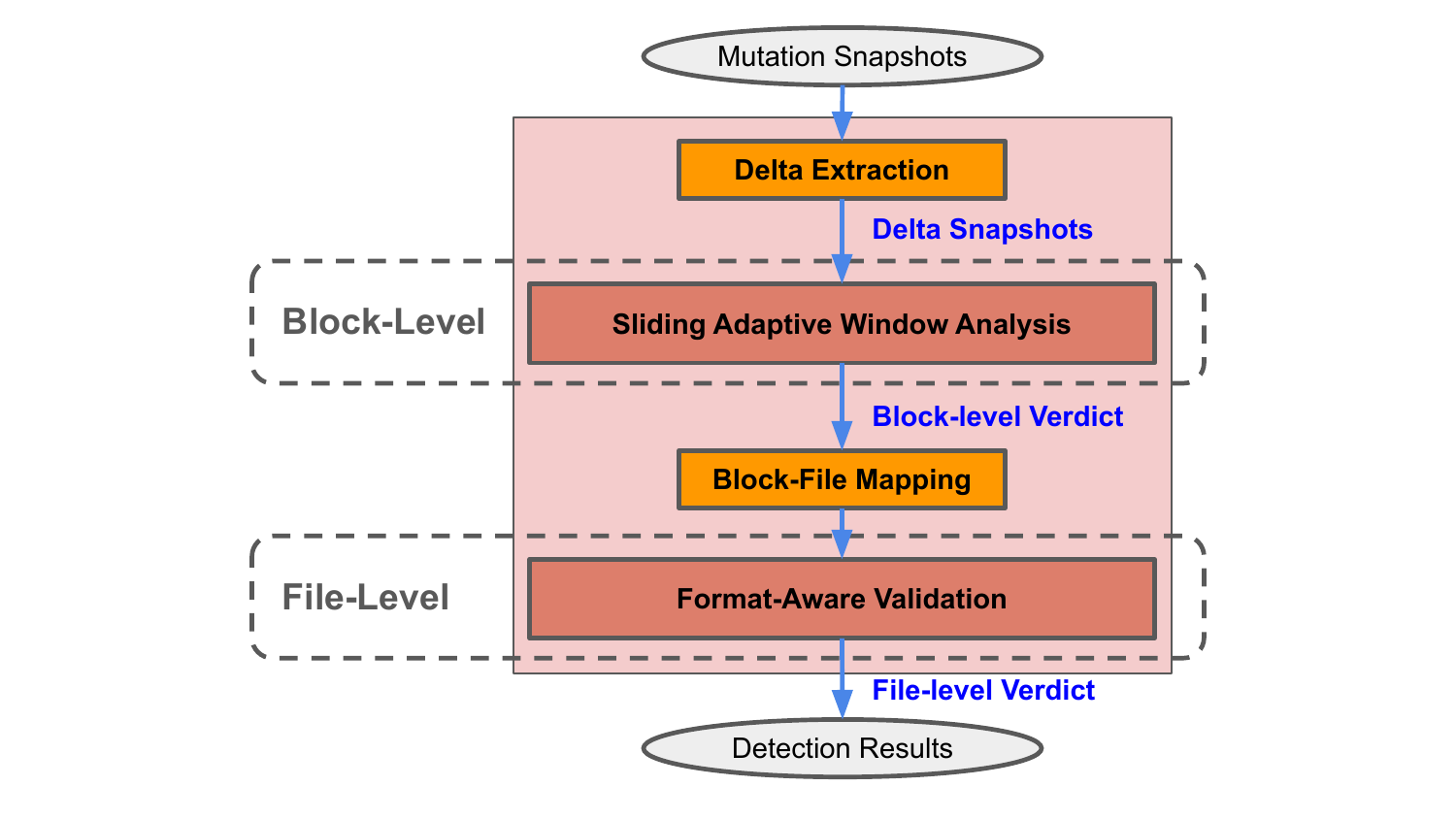}
    \caption{\RansomSaver{}'s Format-Aware Content-Only Detection Pipeline. 
    \RansomSaver{}'s dual-level detection pipeline takes mutation snapshots as inputs and produces detection results as outputs. 
    }
    \vspace{-0.3cm}
    \label{fig:detection_pipeline}
    %\vspace{0.2cm}
\end{figure}
%\FloatBarrier
%\clearpage

\RansomSaver{} \emph{agent} in the cloud consumes mutation snapshots to maintain block-device snapshots and to detect encryption footprints left by ransomware. Figure~\ref{fig:detection_pipeline} illustrates its dual-level detection pipeline, which consists of four stages. While the first two stages employ heuristics to efficiently narrow down candidates, \emph{format-aware validation (FAV)} serves as the authoritative stage that makes the final determination.
\begin{enumerate}
    \item \textbf{Delta extraction:} From mutation snapshots and periodic block-device snapshots, the agent extracts \emph{delta snapshots}, which represent the exact forward differences written to the block device during each epoch. These deltas define the analysis scope for subsequent stages.
    \item \textbf{Sliding Adaptive Window Analysis (SAWA):} SAWA acts as an early prefilter that scans delta snapshots to identify spans for which statistical tests (e.g., $\chi^2$) cannot reliably distinguish plaintext from ciphertext. When encryption is sufficiently coarse-grained, SAWA may already flag blocks as clearly suspicious. However, rather than making definitive decisions, SAWA conservatively forwards \emph{ambiguous or borderline blocks} to the next stage, ensuring high recall.
    \item \textbf{Block–file mapping:}  Blocks flagged by SAWA are mapped back to their corresponding files and converted into file-offset ranges using metadata from block-device snapshots. These ranges localize potentially affected file regions and are passed to the final stage.
    \item \textbf{Format-Aware Validation (FAV):}  FAV is the decisive stage of the pipeline. It inspects suspicious files at the file level, focusing on the file-offset ranges identified earlier, and determines whether the observed data states are invalid. Rather than relying on statistical anomalies, FAV checks whether the data violate the \emph{syntactic or semantic invariants prescribed by the file format specification}. As a result, FAV can accurately detect fine-grained and partial encryption that evades entropy- or $\chi^2$-based methods.
\end{enumerate}

\begin{figure}
    \includegraphics[trim=0cm 2cm 0cm 0cm,width=\linewidth]{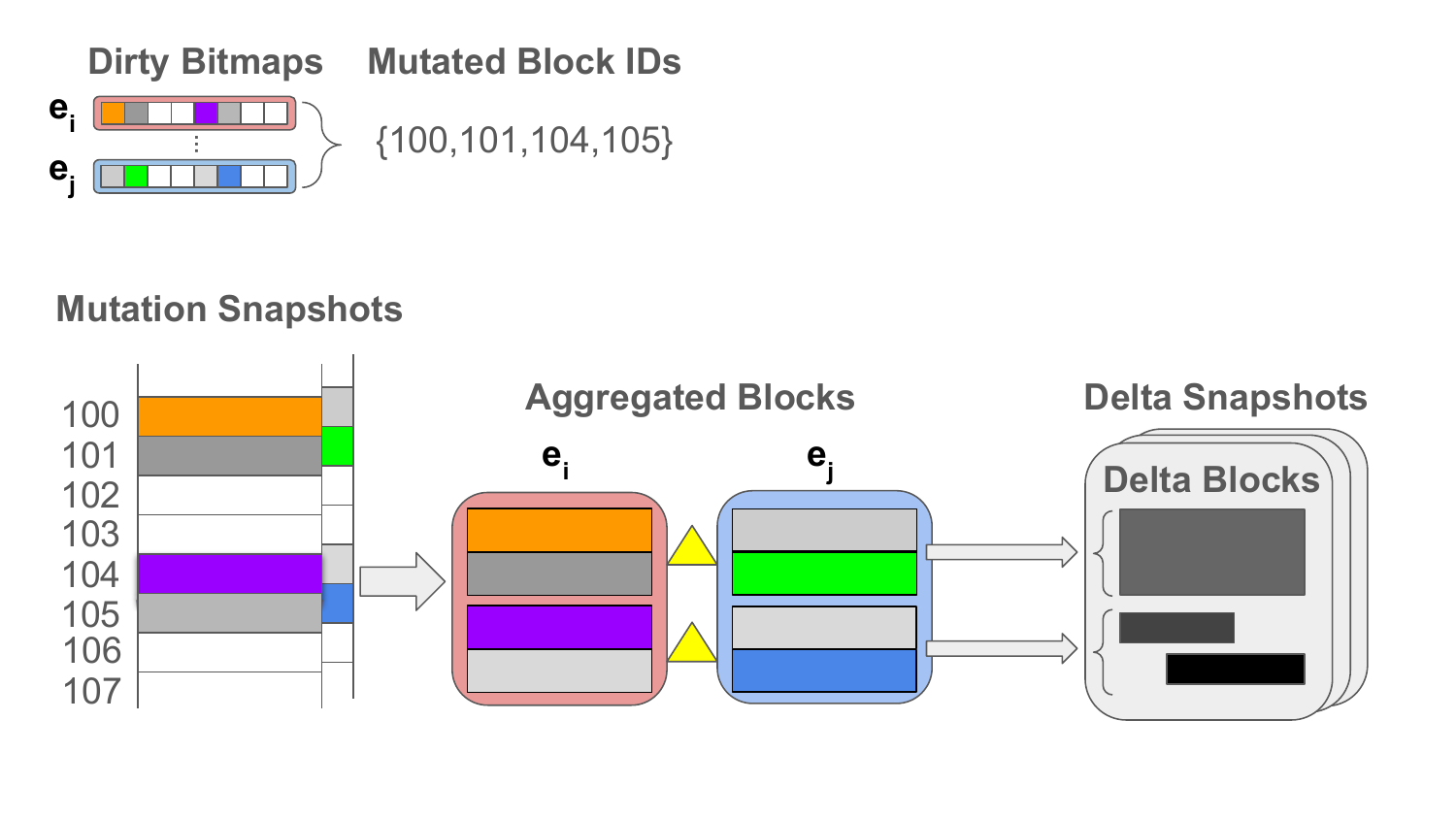}
    \caption{Delta Extraction. 
    \RansomSaver{} computes forward differences of mutated blocks to construct a delta snapshot. 
    }
    \vspace{-0.3cm}
    \label{fig:delta_extract}
    %\vspace{0.2cm}
\end{figure}

\subsection{Delta Extraction}

The delta extraction stage converts raw mutation snapshots into compact \emph{delta blocks} which together constitute a single \emph{delta snapshot} as illustrated in Figure~\ref{fig:delta_extract}. This stage takes two epochs, $e_i$ and $e_j$, where $e_i \leq e_j$, and emits a sequence of delta blocks grouped by extent---an approximate unit of contiguous block allocation in a file system. Each delta block records a contiguous region of modified bytes, that is, it represents the forward difference between the block device snapshot taken before $e_i$ and that taken after $e_j$ for the corresponding extent. This representation is directly consumable by the next stage.

The stage begins by identifying the set of physical block identifiers corresponding to dirty blocks mutated during epochs between $e_i$ and $e_j$, inclusive. Each per-epoch mutation snapshot is accompanied by a dirty bitmap that records which blocks were updated during the given epoch. To determine the complete set of modified blocks within the epoch window $[e_i, e_j]$, \RansomSaver{} performs a logical OR reduction across all dirty bitmaps in that range, effectively computing the union of all mutated blocks.

Next, the collected block identifiers are grouped by the extent to which they belong, where the extent size is fixed at 4{,}096\,B (i.e., an approximation of a file-level extent) and the block size is 512\,B (i.e., an approximation of a disk-level sector). Because the collected block identifiers correspond only to dirty blocks, each extent records only the blocks actually updated within that extent. \RansomSaver{} further coalesces contiguous runs of modified blocks into aggregated blocks per extent. This coalescing approach minimizes short residual spans at block boundaries and allows the subsequent byte-wise differencing stage to process longer, continuous spans more efficiently.

Subsequently, forward differences are computed per aggregated block. For each aggregated block, \RansomSaver{} obtains two byte arrays: one is the latest content before epoch $e_i$ and the other is the latest content from the block snapshots after epoch $e_j$. To compute forward differences, we linearly scan the paired byte arrays to locate modified regions and emit \emph{delta blocks} corresponding to those regions. To mitigate fragmentation due to tiny unchanged islands within a larger write, we treat matching runs shorter than \texttt{min\_gap\_length} (default 16\,B) as part of the surrounding diff. This ``gap suppression'' prevents accidentally splitting a single high-entropy region due to a short run that is coincidentally equivalent between the old and new contents. Finally, all delta blocks are compiled into a \emph{delta snapshot}.

The design is deterministic and embarrassingly parallel: extents are independent units, and aggregated block runs within an extent can be processed sequentially with low memory pressure (streaming over one run at a time) or in parallel when resources permit. Finally, the emitted delta blocks serve as the canonical bridge to downstream stages. Because each delta block corresponds to an exact mutation during the given time window, downstream stages can localize analysis to the modified regions.

\subsection{Sliding Adaptive Window Analysis}\label{subsec:sawa}

% This is an optional prefiltering stage operating at the block-level to select suspicious data state. 
% Using sliding adaptive window analysis, it can relatively accurately pinpoint ciphertext areas within the given chunks of blocks.
% Although this stage may be skipped, the goal here is to filter out blocks that are definitely ciphertext early before file-level checks.
% Since there is no high-level information available in this stage other than byte sequences, statistical methods are utilized in a conservative manner minimizing false negatives. 
% So, SAWA scans the delta snapshots to locate high-entropy writes and flags suspicious blocks, using a $\chi^2$ test with a small initial window size and lenient test score threshold. 
% Blocks that cannot be reliably classified as ciphertext due to the fundamental limitation of a $\chi^2$ test are sent to the downstream for format-aware validation. 
% In this stage, it is possible to use other statistical methods or sampling methods. 
% Note fine-grained partial encryption is not likely to be filtered out in this stage, since SAWA's initial window starts tiny and finds all suspicious small spans with conservatively configured threshold.

\begin{figure}[t]
  \centering
  \includegraphics[trim=1.5cm 6cm 0cm 3cm,width=\linewidth]{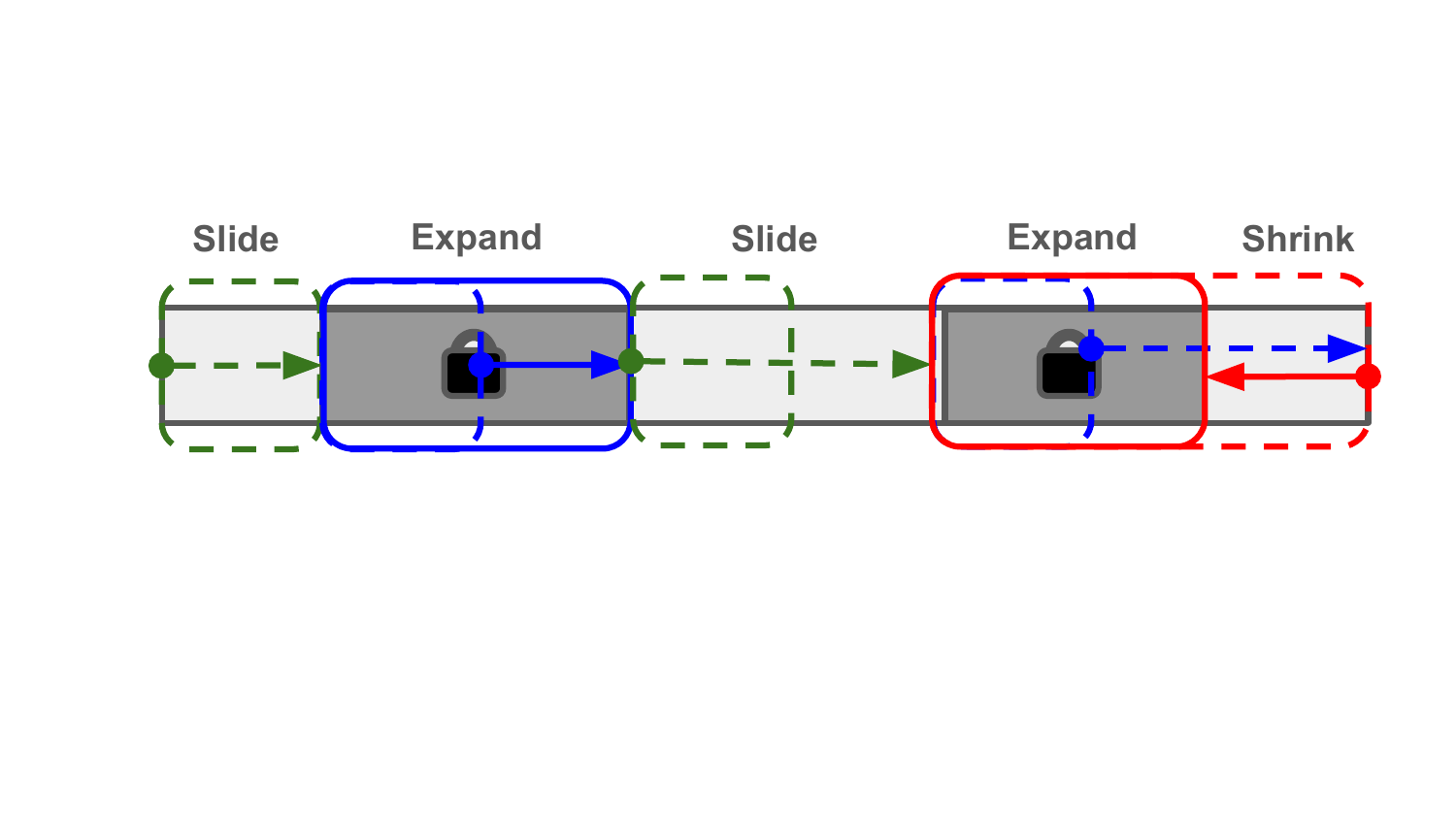}
  \caption{Illustration of the SAWA process.}
  \vspace{-0.3cm}
  \label{fig:sawa-process}
\end{figure}

Figure~\ref{fig:sawa-process} illustrates how the SAWA mechanism operates within our detection pipeline. 
SAWA starts with a fixed initial window width to measure uniformity over a region of the data.
To quantify uniformity, SAWA employs a \emph{chi-squared uniformity test}, which measures the deviation of observed byte frequencies from a perfectly uniform distribution:
\[
\chi^2 = \sum_{b=0}^{255} \frac{(c_b - m/256)^2}{m/256},
\]
where $c_b$ is the observed count of byte value $b$ in a window of length $m$.  
Lower $\chi^2$ values correspond to more uniform, higher-entropy data.  
A window is classified as suspicious when $\chi^2 \le \tau$, where $\tau$ is a configurable threshold.  

SAWA repeats the adaptive cycle of sliding, measuring, expanding, and shrinking to align the detection window with encrypted regions of arbitrary size.
It initially performs a sliding search over each delta block to locate the first offset at which the observed $\chi^{2}$ statistic falls below a predefined uniformity threshold, indicating a potentially encrypted (near-uniform) region.
Upon detecting such a region, the sliding phase pauses, and SAWA transitions to an expansion phase. In this phase, the window size is doubled iteratively to encompass the full suspicious region. 
If the $\chi^{2}$ statistic rises above the threshold (i.e., no longer sufficiently uniform), SAWA goes into the shrinking phase. During the shrinking phase, SAWA employs a binary search to find an optimal boundary, maximizing the window length while maintaining a $\chi^{2}$ statistic below the threshold.
At the end of this process, the expanded window is finalized.
Subsequently, the sliding phase resumes at the end of that finalized window and the expansion phase repeats when another suspicious region is encountered.

\begin{algorithm}[t]
  \caption{Sliding Adaptive Window Analysis}
  \label{alg:sawa}
  \begin{algorithmic}[1]
    \Require List of \texttt{DeltaBlock}s $\mathcal{D}$, window width $w$, stride $s$, uniformity threshold $\tau$, device block size $B$
    \Ensure Set of suspicious block ranges $\mathcal{R}$
    \State $\mathcal{R} \gets \emptyset$
    \ForAll{\texttt{DeltaBlock} $d \in \mathcal{D}$}
        \State $x \gets$ byte payload of $d$
        \State $p \gets 0$
        \While{$p + w \le |x|$}
            \State $L \gets w$
            \State $chi \gets \chi^2(x[p{:}p{+}L])$
            \If{$chi \le \tau$}
                \While{$chi \le \tau$ \textbf{and} $p+2L \le |x|$} 
                    \State $L \gets 2 \times L$ \Comment{geometric expansion}
                    \State $chi \gets \chi^2(x[p{:}p{+}L])$ 
                \EndWhile
                \State Maximize window length $L$ \textbf{s.t.} $chi \le \tau$
                \State $range \gets $ Get a Block Range for $[p{:}p{+}L)$
                \State $\mathcal{R} \gets \mathcal{R} \cup \{range\}$
                \State $p \gets p + L$
            \Else
                \State $p \gets p + s$
            \EndIf
        \EndWhile
    \EndFor
    \State \Return $\mathcal{R}$
  \end{algorithmic}
\end{algorithm}

Algorithm~\ref{alg:sawa} illustrates the overall workflow.  
The procedure begins with a forward scan using a sliding base window of width $w$ (e.g., 16\,B) and stride $s$ (e.g., 4\,B).  
When a window passes the uniformity test, SAWA anchors that position and enters an adaptive expansion phase, repeatedly doubling the window size as long as uniformity holds.  
This geometric growth avoids fragmentation of a single long high-entropy region into small fragments, stabilizing the reliability of the uniformity measurement.  
A shrinking phase may follow to fine-tune the end of the window boundary, aiming to yeild the nearly maximal span that remains below the threshold.

The SAWA stage is designed to isolate localized byte ranges within each delta block that display near-uniform byte frequencies. Because statistical methods such as $\chi^{2}$ tests are unreliable for small window widths, this stage serves as a heuristic for identifying potential signs of high entropy often produced by compression or encryption.  SAWA emits records of suspicious regions per delta block, each of which precisely retains physical provenance information, including covered device block identifiers and offsets within corresponding blocks. With this information, subsequent format-aware analyzers can inspect only the relevant byte spans and produce the final decision on whether a PEER attack has occurred. The output of SAWA is intentionally conservative---SAWA serves as a high-recall, locality-preserving prefilter for later stages.  

\subsection{Block-to-File Mapping}

The block-to-file mapping stage translates the suspicious regions, identified by SAWA at the physical block level, into file paths and byte offsets.
\RansomSaver{} scans the file-system metadata to reconstruct the layout of each file’s data regions, represented as a set of non-overlapping intervals in the physical block address space.
The mapping from block identifiers to file paths and offsets is maintained in an in-memory, interval-tree-based index that supports efficient lookups.

Given a set of suspicious block identifiers, the mapper queries the interval index to obtain the corresponding file identifiers and canonical paths.
Because file fragmentation is common, multiple disjoint regions may belong to the same file; the mapper deduplicates these to produce a minimal list of suspicious files that collectively cover all flagged blocks.
This enables the system to localize analysis to the precise file path–offset pairs associated with the suspicious regions reported by SAWA, with minimal overhead.

\scbf{Trusted Manifest and Whitelisting} To reduce false positives, the detection pipeline applies a \emph{whitelisting filter} that verifies the integrity of media and other static files against a curated whitelist. The core assumption is that such files (e.g., images, videos, and other published media assets) are rarely modified after distribution, and can therefore be safely trusted if their integrity remains intact. Therefore, the detection pipeline delegates the verification of file formats such as \texttt{.jpg}, \texttt{.png}, \texttt{.mp3}, and \texttt{.mp4} to the whitelisting filter. 

After block-to-file mapping resolves suspicious blocks to specific files, the pipeline cross-checks each candidate file against a manifest of trusted content. The manifest stores cryptographic hash values (SHA-256) of files previously published by trusted sources. For each suspicious file, the pipeline computes its current hash and compares it to the manifest entry. Files whose hashes match a trusted record are marked as verified and are subsequently excluded from further analysis.

This mechanism ensures that benign but frequently accessed content—such as cached media or bundled application assets—is not repeatedly flagged as suspicious. By validating integrity rather than relying solely on file names or paths, the filter provides strong guarantees against both accidental and adversarial false positives. The resulting set of suspicious files not whitelisted is passed downstream to the format-aware filter for in-depth inspection.

\subsection{Format-Aware Validation}

Once candidate suspicious byte ranges are identified—either by SAWA (for blocks at or above the minimum $X^2$ window size) or by directly forwarding smaller blocks that cannot be reliably scored—the pipeline performs \emph{format-aware validation} to refine these preliminary results using format-specific syntax and semantics. Rather than treating files as opaque byte sequences, the validator reconstructs their logical content—including decompression or other decoding steps—and evaluates whether the decoded structure conforms to the syntactic and semantic invariants of the declared file type. This stage dramatically reduces false positives while preserving encrypted or corrupted regions that violate format expectations.
Each file containing candidate regions is dispatched to a specialized validator based on its magic bytes and extension (e.g., \texttt{.txt}, \texttt{.zip}, \texttt{.xlsx}, \texttt{.pptx}, \texttt{.docx}, \texttt{.pdf}). Validators incorporate deep, format-aware reasoning—parsing container metadata, inflating compressed streams, detecting encrypted members, and interpreting content-specific structures—to determine whether a flagged region truly represents ciphertext, benign high-entropy data (e.g., images, fonts, media), or small plaintext spans.
By analyzing decoded and interpreted content rather than raw bytes, format-aware validators ensure that only genuinely anomalous, semantically implausible, or unverifiable regions remain escalated.

\scbf{Text Validator}
The text validator targets formats that are expected to contain human-readable content, such as logs, configuration files, and source code. Its core intuition is that textual data is encoded using highly structured schemes such as UTF-8, whereas ciphertext behaves like random bytes and almost never conforms to such syntactic constraints. The text validator therefore applies strict UTF-8 decoding to each suspicious span. Genuine plaintext must decode cleanly, while ciphertext or corrupted bytes are highly likely to produce decoding failures. Successfully decoded spans are dismissed as benign, but decoding failures cause the region to be escalated as suspicious ciphertext.

Although an attacker could, in principle, generate ciphertext that satisfies UTF-8’s syntactic constraints, producing ciphertext that also exhibits semantic coherence in natural language is computationally infeasible under standard cryptographic assumptions, as ciphertext is designed to be indistinguishable from random. That is, even when random ciphertext accidentally decodes as UTF-8, it almost never exhibits semantic coherence in natural language, which can be readily detected using analysis incorporating natural language processing (NLP).

\scbf{ZIP Validator}
The ZIP validator distinguishes legitimate compressed content from true ciphertext by interpreting the archive’s container structure rather than treating the archive as a flat byte sequence.
It first recovers the boundaries of the \emph{end-of-central-directory (EOCD)}, the \emph{central directory (CD)}, and local file headers, then partitions the file into three logical regions: (i) metadata (EOCD/CD), (ii) member payloads, and (iii) unclaimed gaps.

Metadata regions receive structural validation to ensure header and record correctness.
The validator checks field consistency across EOCD records, CD entries, and local headers, verifying offsets, flags, lengths, and signature patterns.
%When the data-descriptor flag is set, the validator additionally validates the presence and consistency of the data descriptor (CRC and sizes) against the corresponding central-directory entry.
These checks distinguish legitimate structural bytes from anomalous uniform-looking or malformed segments, which are escalated.

Member payloads are interpreted according to their compression method.
\texttt{DEFLATE} members are inflated, while \texttt{STORE} members are read verbatim.
%For each decoded member payload, the validator additionally verifies integrity by checking CRC-32 consistency with the archive metadata.
The ZIP validator recursively descends into nested containers (e.g., ZIP-within-ZIP) until decoded \emph{leaves} are reached or depth and byte budgets are exhausted.
All classification is performed on decoded bytes, preventing compressed high-entropy data from being mistaken for ciphertext.
Encrypted members and media leaves (e.g., JPEG, PNG, MP3, MP4) bypass decoding-based analysis and are instead checked against a trusted manifest.
Encrypted members are validated using their encrypted payload hashes, while media leaves are validated using content hashes computed over either decoded or raw bytes.
Items with matching hashes are whitelisted.
Text-like leaves undergo plaintext checks using the text validator's functionality described earlier.

Unclaimed gaps arise from padding, alignment, partially written archives, or malformed ZIPs. Since attackers may hide ciphertext in such slack space, the validator applies size- and entropy-aware guards: small, low-entropy paddings are tolerated, while unusually large, high-entropy, or structurally implausible gaps are conservatively escalated as potential injected ciphertext or hidden malicious content.

%Unclaimed gaps can arise from padding, alignment, partially written archives, or malformed ZIP files. The ZIP validator treats such gaps as malicious, as attackers may hide ciphertext in this slack space.

\scbf{OOXML Validator}
OOXML formats (\texttt{.docx}, \texttt{.xlsx}, \texttt{.pptx}) are ZIP-based containers that organize content into structured parts, such as XML files for document text, slides, or worksheets; embedded media files; and metadata files (e.g., styles, relationships, themes) that describe how the parts connect. The OOXML validator follows the same member-first, decode-then-decide approach as the ZIP validator: it enumerates all members via local headers and, for each suspicious SAWA span, promotes the classification decision derived from the overlapping member rather than from mid-stream bytes.

Member payloads are obtained by decoding the underlying container (\texttt{DEFLATE} inflation or \texttt{STORE} passthrough), ensuring that all classification is performed on \emph{decoded} content rather than on raw compressed streams. Media parts (e.g., \texttt{word/media/}, \texttt{ppt/media/}, \texttt{xl/media/}, and thumbnails) are validated through a trusted manifest: matching SHA-256 digests are whitelisted, while hash mismatches or unreadable media are escalated as suspicious.

Text-centric parts, including \texttt{word/document.xml}, slide XML, worksheet XML, and shared-strings tables, undergo plaintext validation via strict UTF-8 decoding and lightweight XML sanity checks. Malformed XML, mixed binary/XML regions, or implausible textual structure lead to escalation, while semantically consistent text-like content is dismissed as benign.

Suspicious spans not overlapping any member are interpreted as metadata regions (local-file-header (LFH) ranges, data-descriptor trailers, or small pre-LFH preambles). These regions receive conservative structural validation, and only uniform-like or anomalously large gaps inconsistent with the OOXML container layout are escalated, preventing spurious alerts on expected structural bytes.

\scbf{PDF Validator}
The PDF validator is stream-aware: it maps byte ranges to concrete PDF
streams and their adjacent structural metadata (e.g., headers,
\texttt{xref}, \texttt{trailer}, and \texttt{startxref}), enabling
per-stream decisions rather than coarse, file-wide heuristics.
Lightweight dictionary parsing extracts \texttt{/Filter} and
\texttt{/DecodeParms} entries, allowing the validator to decode supported
filter pipelines (Flate, ASCII85, ASCIIHex, RunLength, and PNG predictors
for Flate).
Whenever decoding is possible, validation operates on the decoded
representation of a stream, ensuring that decisions reflect its logical
structure rather than its compressed or encoded form.

Certain stream types, most notably image and font streams, contain opaque
binary payloads whose internal structure is only weakly constrained by the
PDF specification. For such streams, the validator applies a
whitelist-based integrity check: a cryptographic hash (SHA-256) is
computed over a canonical representation of the stream payload and
compared against a trusted manifest provided by the PDF publisher.
Other component types such as ICC profiles, content streams, object
streams, embedded files, and XMP metadata are validated by checking
whether their decoded bytes exhibit the minimal structural
characteristics expected for that object class (e.g., recognizable
dictionary prefixes, valid stream headers, or well-formed metadata
markers). Components whose decoded representations appear malformed,
inconsistent with their declared type, or unnaturally uniform are
escalated as potential ciphertext.
Regions outside any stream including the PDF header, object dictionaries,
\texttt{xref} tables, trailers, and small gaps introduced by alignment or
incremental updates are validated using simple structural checks rather
than content analysis. These regions typically consist of short metadata
fragments with predictable markers (e.g., ``\%PDF-1.x'', ``xref'',
``trailer'', object numbers, and dictionary delimiters).

\section{Implementation Details}\label{sec:implement}

\scbf{\RansomSaver{} Frontend and Backend} \RansomSaver{} is built on \emph{Rocky}~\cite{rocky}, an open-source distributed replicated block device designed to be a failure- and tamper-resistant storage for edge cloud environments. \emph{Rocky} implements the design of \RansomSaver{} \emph{frontend} and \RansomSaver{} \emph{backend} described earlier in Section~\ref{subsec:design_arch}. For convenience, \emph{Rocky endpoint} and \emph{connector-cloudlet} are renamed as \RansomSaver{} \emph{frontend} and \RansomSaver{} \emph{backend}, respectively. The implementation details of those are discussed in the original paper~\cite{rocky}.

\scbf{\RansomSaver{} Agent} We implemented the prototype of \RansomSaver{} Agent in Python and the main detection pipeline on top of \emph{Apache Spark} to enable scalable, fault-tolerant processing of large snapshot datasets. The entire implementation consists of approximately 10K lines of Python code organized into three components~\footnote{We report lines-of-code (LoC) for the three core pipeline components using \texttt{cloc}.}: preprocessor (292 LoC), detector (7{,}341), and postprocessor (655 LoC). Preprocessor fetches mutation snapshots and bitmaps from the \RansomSaver{} \textit{backend} to locally persist those and construct block device snapshots for each epoch. Detector runs the detection pipeline described earlier. It contains independent subsystems that mirror the logical stages of detection: delta extraction, SAWA, block-to-file mapping, and FAV. Within the detector, FAV (\texttt{detector/fileaware/}) contributes 4{,}527 LoC, while the remaining detector logic (delta extraction, SAWA, block-to-file mapping, orchestration) accounts for 2{,}814 LoC. In sum, FAV constitutes the majority of the detector (\(\approx\)62\% of detector code). The detector is modular and highly extensible, therefore new format-aware validators can be added  for unsupported file formats or more powerful format-aware filters can be added.

%The trained neural network had 1,227,777 parameters, trained with a batch size of 64 for 70 steps taking one minute and thirty seconds. We used ResNet as a base model with a sigmoid activation function for the final layer and rectified linear activation for other layers. We used Binary Cross Entropy (BCELoss) to calculate the loss and optimized using Adam with a learning rate of 0.001.

% \scbf{Time-Traveling Support}
% To enable mounting block-device snapshots in read-only mode for any given epoch, \RansomSaver{} integrates a time-traveling mechanism built atop \textit{Rocky}. The implementation consists of 309 lines of Java code for managing historical state transitions and 70 lines of Proto3 definitions for the gRPC interface. The system supports two core primitives: \texttt{Rewind} and \texttt{Replay}. \texttt{Rewind} resets the device to the initial state and reapplies forward deltas up to the target epoch, ensuring consistency regardless of the current position in time. In contrast, \texttt{Replay} continues from the current epoch and incrementally applies forward deltas to reach the requested future state. For greater efficiency, future versions may support placing auxiliary snapshots at intermediate epochs or incorporating inverse deltas to reduce replay latency. Together, these primitives allow \RansomSaver{} to navigate storage history, supporting retrospective analysis.

\section{Evaluation}\label{sec:eval}

Our evaluation aims to answer the following research questions:
\begin{enumerate}
    \item How effectively can \RansomSaver{} detect PEER attacks under diverse partial-encryption strategies, and what is the performance profile of the entire detection pipeline?
    \item Does \RansomSaver{} more effectively detect PEER attacks using fine-grained partial-encryption strategies than previously proposed content-based detection methods based on statistical analysis?
    \item Can \RansomSaver{} detect known attacks from real-world evasive ransomware variants?
\end{enumerate}

\subsection{Detection of Various Evasive Attacks}\label{subsec:detect_various}

We evaluate the effectiveness and efficiency of \RansomSaver{} in detecting various evasive attacks by mounting simulated attacks and detecting those with full detection pipeline described earlier. 

\scbf{Environment Setup}
%Our experimentation was conducted on 6 desktop PCs in our lab. Desktop PCs are equipped with an Intel i5 processor, 16\,GB RAM, and 512\,GB NVMe SSD. The host OS was Ubuntu 22.04 on all of them. Among 6, five desktop PCs are used to form a Spark cluster. On one of non-Spark machine, we run a VirtualBox VM installed with Windows 10 Home Edition. Then, the VM is configured to have a virtual block device as the boot partition and a second virtual block device backed by the \RansomSaver{} \emph{frontend}. The size of the virtual block device for boot partition is set to 50\,GB, while the size of the virtual block device backed by the \RansomSaver{} \emph{frontend} is 1\,GB---it is not an upper bound but just minimized for the evaluation purpose. In addition, the VM serves as a contained environment to run ransomware samples. Meanwhile, an instance of \RansomSaver{} backend run on one of the Spark machine. 
Our experiments were conducted on a cluster of five desktop PCs in our lab. 
Each machine is equipped with an Intel i5 processor, 16\,GB of RAM, and a 512\,GB NVMe SSD, and runs Ubuntu~22.04 as the host operating system. The five machines are configured as a Spark cluster.
The \RansomSaver{} frontend, backend, and agent are co-located on a single machine within the cluster. Workload generation, preprocessing, and detection are executed as separate stages. Because these stages are run independently, co-locating components does not affect detection effectiveness or performance measurements.

To evaluate the detection capability of \RansomSaver{}, we collected one representative file for each of ten file formats, totalling approximately 16.4\,MB. The set spans common desktop formats and containerized documents: a PDF (105{,}116\,B); OOXML documents including PowerPoint (59{,}960\,B), Excel (12{,}817\,B), and Word (75{,}973\,B); image files including JPEG (142{,}294\,B) and PNG (71{,}291\,B); a ZIP archive (64{,}829\,B); audio and video media such as MP3 (480{,}967\,B) and MP4 (16{,}172{,}912\,B), respectively; and a plain text file (28{,}324\,B). 

On the detection side, we fix the device block size to 512\,B and the extent size to 4\,KB, and we use a single set of SAWA parameters across all runs. In particular, we set the window width to be $w{=}16$\,B and the stride to be $s{=}8$\,B, resulting in 50\% overlap between adjacent window measurements. For the statistical test, each window is scored by a $\chi^2$ test using a 256-bin byte-uniform model (df $=255$), as described in Section~\ref{subsec:sawa}. We use a $\chi^2$ threshold of $\tau{=}350$ to separate high-entropy segments from benign content. The chosen combination of parameters, ($w{=}16$\,B, $s{=}8$\,B, $\tau{=}350$), was selected from pilot sweeps to balance false-positive control with strong recall under PEER attack patterns described below.

\begin{table}[t]
\centering
\caption{Simulated PEER Attack Patterns for Evaluation}
\label{tab:patterns}
\renewcommand{\arraystretch}{1.2}
\small
\begin{tabular}{p{2.8cm} p{5.0cm}}
\toprule
\textbf{Pattern (\textit{Parameters})} & \textbf{Definition} \\
\midrule
% Full ()
% & Encrypts the entire file sequentially from offset 0 to EOF. \\
% \midrule
Fast ($N$) 
& Encrypts only the first $N$ bytes (prefix) of the file. \\
\midrule
SkipStep ($N, S$) 
& Repeats a stripe pattern: encrypts $N$ bytes, then skips $S$ bytes, continuing until EOF. \\
\midrule
% Smart ($N, P$) 
% & Encrypts the first $N$ bytes, then at each $\tfrac{1}{P}$ fraction of the file length, encrypt the next $N$ bytes. \\
% \midrule
Animagus ($F$) 
& Randomly encrypts approximately $F\%$ of file blocks. \\
% \midrule
% DeviceSpray ($T, N, R, D$) 
% & Encrypts $T$ bursts of length $N$ each, written at pseudo-random block offsets within the range $R = [B_{\min}, B_{\max}]$ on the device, bypassing the filesystem namespace. The PRNG is controlled by seed $D$. \\
\bottomrule
\end{tabular}
\end{table}

\scbf{Attack Scenarios}
%Our PEER simulator clones original files and encrypts them using a 256-bit AES block cipher in CTR mode according to five attack patterns as described in Table~\ref{tab:patterns}: \emph{Full}, \emph{Fast}, \emph{Skip-Step}, \emph{Smart}, and \emph{Animagus}. These attack patterns are drawn from behaviors of traditional ransomware that encrypts entire files~\cite{acronis_ransomware_overview}, existing evasive ransomware that employs partial encryption---such as Black Basta~\cite{SRLabs2024BlackBastaBuster}, BlackCat~\cite{ransomware-blackcat-rootkit-2023}, and Play~\cite{ransomware-play-2023}~\footnote{We do not claim that these patterns are exhaustive.}---and research prototype ransomware that conducts imitation attacks~\cite{animagus}. \emph{Full} encryption encrypts the entire file and is maximally destructive but inefficient, reflecting largely traditional ransomware behavior. \emph{Fast} encryption targets headers and indexes to break usability with minimal runtime. \emph{Skip-Step} alternates encrypted and plaintext regions, creating partial corruption that can bypass na\"{\i}ve detectors employing statistical analysis. \emph{Smart} distributes damage by combining a header prefix with partial coverage in subsequent blocks. \emph{Animagus} randomly selects a fraction of file blocks to encrypt and mixes encrypted blocks with plaintext within each write buffer, reducing per-write entropy while still rendering files unrecoverable. 
Our PEER simulator clones original files and encrypts them using a 256-bit AES block cipher in CTR mode according to three attack patterns as described in Table~\ref{tab:patterns}: \emph{Fast}, \emph{Skip-Step}, and \emph{Animagus}. These attack patterns are drawn from behaviors of existing evasive ransomware that employs partial encryption---such as Black Basta~\cite{SRLabs2024BlackBastaBuster}, BlackCat~\cite{ransomware-blackcat-rootkit-2023}, and Play~\cite{ransomware-play-2023}~\footnote{We do not claim that these patterns are exhaustive.}---and research prototype ransomware that conducts imitation attacks~\cite{animagus}. \emph{Fast} encryption targets headers and indexes to break usability with minimal runtime. \emph{Skip-Step} alternates encrypted and plaintext regions, creating partial corruption that can bypass na\"{\i}ve detectors employing statistical analysis. \emph{Animagus} randomly selects a fraction of file blocks to encrypt and mixes encrypted blocks with plaintext within each write buffer, reducing per-write entropy while still rendering files unrecoverable. 

%For each simulated PEER attack campaign, we execute the corresponding attack pattern and run the \RansomSaver{} agent to measure the detection accuracy. Initially, we format the block device exposed by the \RansomSaver{} frontend to install the NTFS file system using the \texttt{mkfs.ntfs} command. Then, we mount the block device and copy all original files into the mount point. Subsequently, the PEER simulator clones and encrypts files following the specific attack pattern configured for the run.  For the \emph{Fast}, \emph{Skip-Step}, and \emph{Smart} patterns, we set the ciphertext chunk size to 4\,KB. For \emph{Skip-Step}, the simulator skips 4\,KB before encrypting the next chunk. For \emph{Smart}, we set the fraction of the file length used for encryption to 30\%. For \emph{Animagus}, we configure the simulator to encrypt 25\% of file blocks.
For each simulated PEER attack campaign, we execute the corresponding attack pattern and run the \RansomSaver{} agent to measure the detection accuracy. Initially, we format the block device exposed by the \RansomSaver{} frontend to install the NTFS file system using the \texttt{mkfs.ntfs} command. Then, we mount the block device and copy all original files into the mount point. Subsequently, the PEER simulator clones and encrypts files following the specific attack pattern configured for the run.  For the \emph{Fast} and \emph{Skip-Step} patterns, we set the ciphertext chunk size to 4\,KB. For \emph{Skip-Step}, the simulator skips 4\,KB before encrypting the next chunk. For \emph{Animagus}, we configure the simulator to encrypt 25\% of file blocks.

\begin{table*}[ht]
\centering
\caption{\textbf{Block-level and file-level results per simulated PEER attack pattern for sample data.} 
}
\label{tab:persim-sample-data-result}
%\small
\setlength{\tabcolsep}{6pt}
\begin{tabular}{lrrrrrrrrrrrrrr}
\toprule
\multirow{2}{*}{Attack Pattern} &
\multicolumn{4}{c}{Block-Level} & 
\multicolumn{4}{c}{File-Level} &
\multicolumn{4}{c}{Performance Metrics} \\
\cmidrule(lr){2-5}\cmidrule(lr){6-9}\cmidrule(lr){10-13}
& TP & TN & FP & FN & TP & TN & FP & FN & Accuracy & Precision & Recall & F1-Score\\
\midrule
Fast    &    80 & 34379 & 32957 &  0 & 10 & 12 & 0 & 0 & \multirow{3}{*}{\makecell[l]{100\%}} & \multirow{3}{*}{\makecell[l]{100\%}} & \multirow{3}{*}{\makecell[l]{100\%}} & \multirow{3}{*}{\makecell[l]{100\%}}\\
Skip    & 16837 & 34027 & 16552 &  0 & 10 & 12 & 0 & 0 & \\
Animagus    &  8394 & 34328 & 24694 & 0 & 10 & 12 & 0 & 0 & \\
\bottomrule
\end{tabular}
\vspace{4pt}
\end{table*}

\scbf{Detection Results}
Table~\ref{tab:persim-sample-data-result} shows the results of the described experiment.
We compute a confusion matrix at the block level to evaluate the performance of the SAWA-based analysis.
Subsequently, we compute a confusion matrix at the file level to evaluate the performance of the file-level FAV-based analysis.
At the block level, recall is effectively perfect for most attack patterns (FN $=0$), while precision varies widely depending on the mutation strategy.
Evasive attack patterns such as \emph{Fast}, \emph{Skip-Step}, and \emph{Animagus} generate many ciphertext-like regions, which depress block-level precision despite high recall.
Nevertheless, at the \emph{file level}, \RansomSaver{} combines suspicious blocks on a per-file basis and applies file-aware validators, eliminating spurious alarms across all settings.
As a result, the file-level analysis achieves 100\% accuracy, precision, recall, and F1 score.

\begin{table*}[ht]
\centering
\caption{Performance cost per attack pattern (times in seconds) with written/encrypted data.}
\label{tab:perf-costs}
\begin{tabular}{lrrrrrccrr}
\toprule
\multirow{2}{*}{Attack Pattern}
  & \multicolumn{5}{c}{Time (s)}
  & \multirow{2}{*}{Total}
  & \multirow{2}{*}{Written}
  & \multicolumn{2}{c}{Encrypted}\\
\cmidrule(lr){2-6}\cmidrule(lr){9-10}
& Delta & SAWA & Mapping & FAV & Output
& (mean $\pm$ std)
& Blocks & Blocks & Bytes\\
\midrule
Fast           & 5.486 & 10.082 & 1.193 & 6.101 & 2.395 & 25.257 $\pm$ 0.430 & 67416 & 80    & 40960    \\
Skip           & 5.727 & 10.318 & 1.203 & 8.840 & 2.372 & 28.460 $\pm$ 0.657 & 67416 & 16837 & 8619239  \\
Animagus       & 6.077 & 11.643 & 1.155 & 6.020 & 2.532 & 27.427 $\pm$ 0.570 & 67416 & 8394  & 4297728  \\
\bottomrule
\end{tabular}
\end{table*}

%\subsection{Performance Overheads}

\scbf{Performance Profile}
% Table~\ref{tab:perf-costs} summarizes the end-to-end runtime across five attack patterns. We run each attack pattern and detection three times and compute the average as well as standard deviation (std.). Overall average time taken is 29.45\,s with standard deviation 3.01\,s. The fastest configurations is \emph{Fast} (25.257\,s) and then \emph{Animagus} (27.427\,s). \emph{Skip} (28.460\,s) and then \emph{Smart} (33.579\,s) follow. Lastly, \emph{Full} (33.518\,s) are the slowest. Decomposing the totals by logical stage, the dominant contributors on average are \textsc{SAWA} (12.28\,s) and \textsc{FAV} (7.03\,s). \textsc{Delta} contributes a steady (5.97\,s), while \textsc{Mapping} (1.30\,s) and \textsc{Output} (2.43\,s) remain minor. Variability is likewise concentrated in \textsc{SAWA} (std.\ 1.66\,s) and \textsc{FAV} (std.\ 1.31\,s), whereas \textsc{Mapping} and \textsc{Output} are fast and stable.
Table~\ref{tab:perf-costs} summarizes the end-to-end runtime across the three attack patterns. For each pattern, we run the attack and detection pipeline three times and report the average runtime along with the standard deviation (std.). The overall average runtime is 29.45\,s with a standard deviation of 3.01\,s. The runtime is proportional to the number of blocks encrypted. The fastest configuration is \emph{Fast} (25.26\,s), as it encrypts the smallest number of blocks---approximately 80 blocks, totaling about 40\,KB. The \emph{Animagus} pattern takes 27.43\,s and encrypts approximately 8\,K blocks (about 4.1\,MB). The slowest configuration is \emph{Skip-Step}, which takes 28.46\,s and encrypts approximately 17\,K blocks (about 8.2\,MB). 

Decomposing the total runtime by logical stage, the dominant contributors on average are \textsc{SAWA} (12.28\,s) and \textsc{FAV} (7.03\,s). The \textsc{SAWA} cost remains relatively stable across all patterns because the dataset contains many benign high-entropy files, which require a similar amount of block-level processing regardless of the attack pattern. The \textsc{FAV} cost remains comparable between the \emph{Fast} and \emph{Animagus} patterns because most encrypted regions overlap with compressed regions. As a result, these regions must be decompressed regardless of whether encryption occurs, leading to similar processing costs. However, when encryption increasingly affects non-compressed regions---as in the \emph{Skip-Step} pattern---the \textsc{FAV} cost rises. In particular, the \textsc{FAV} runtime increases by approximately 47\%, as \emph{Skip-Step} encrypts roughly twice as many regions as \emph{Animagus}.

\begin{figure*}[t]
    \centering

    % Row 1
    \begin{subfigure}[t]{0.45\textwidth}
        \centering
        \includegraphics[width=\linewidth]{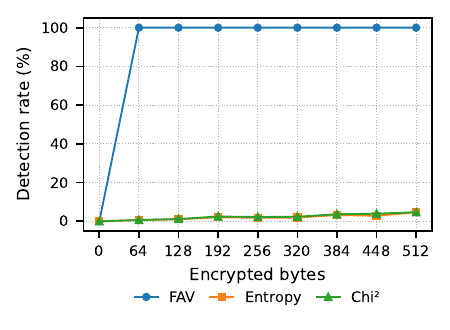}
        \vspace{-0.7cm}
        \caption{PDF files.}
        \label{fig:content-cmpr:pdf}
    \end{subfigure}
    %\hfill
    \begin{subfigure}[t]{0.45\textwidth}
        \centering
        \includegraphics[width=\linewidth]{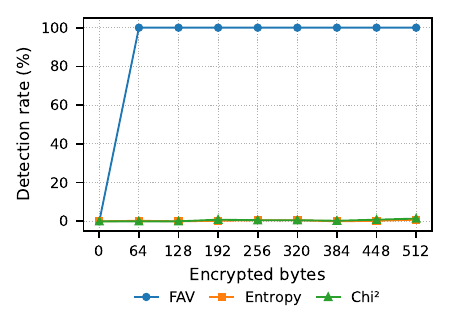}
        \vspace{-0.7cm}
        \caption{DOCX files (OOXML files).}
        \label{fig:content-cmpr:docx}
    \end{subfigure}

    %\vspace{2mm}
    \vspace{1mm}

    % Row 2
    \begin{subfigure}[t]{0.45\textwidth}
        \centering
        \includegraphics[width=\linewidth]{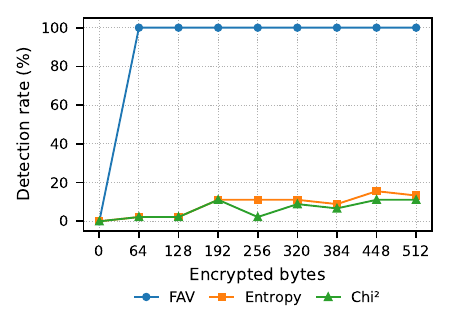}
        \vspace{-0.7cm}
        \caption{ZIP files.}
        \label{fig:content-cmpr:zip}
    \end{subfigure}
    %\hfill
    \begin{subfigure}[t]{0.45\textwidth}
        \centering
        \includegraphics[width=\linewidth]{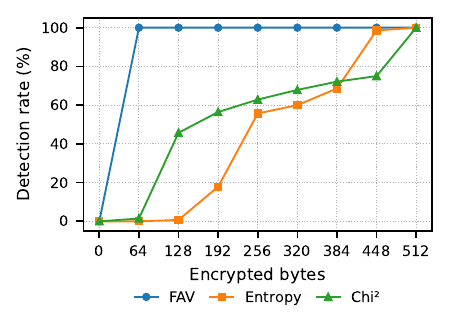}
        \vspace{-0.7cm}
        \caption{TXT files.}
        \label{fig:content-cmpr:txt}
    \end{subfigure}

    \caption{\textbf{Comparison with content-based detection under fine-grained partial encryption.} 
    Detection rate as a function of encrypted bytes per write under a fixed skip distance (skip=128) across four file types.
    \textsc{FAV} (\RansomSaver{}) remains robust under sparse encryption, while statistical detectors degrade as encryption becomes small and scattered.
    }
    \label{fig:content_based_comparison}
\end{figure*}

\subsection{Detection of Fine-Grained Attacks}\label{subsec:detect_fine}

To demonstrate that \RansomSaver{} detects fine-grained partial-encryption attacks more effectively than prior content-based detectors relying on statistical methods, we implement pure file-level detectors based on Shannon entropy and the $\chi^2$ test, and FAV---without SAWA---and compare their detection rates.

\scbf{Environment Setup}
Our experiments for detecting fine-grained attacks were conducted on a single PC in our lab, whose configuration is described earlier in Section~\ref{subsec:detect_various}.
To evaluate the detection capability of \RansomSaver{} against fine-grained file encryption, we collected 140 benign text files (321\,MB), 45 benign ZIP files (442\,MB), 353 DOCX files (325\,MB), and 765 PDF files (651\,MB).
For the entropy- and $\chi^2$-based detectors, thresholds are tuned to achieve the highest sensitivity that incurs no false positives when no encryption is applied.

\scbf{Attack Scenarios}
We simulate fine-grained partial-encryption attacks to stress-test our FAV-based detection beyond existing real-world ransomware attacks. Using the \textsc{PEER} simulator, we clone files and apply a \emph{Skip-Step} encryption pattern.
The number of encrypted bytes per write is varied from 64\,B to 512\,B, while the skip distance is fixed at 128\,B, following the attack pattern observed in Black Basta ransomware variants~\cite{SRLabs2024BlackBastaBuster}.

\scbf{Detection Results}
Figure~\ref{fig:content_based_comparison} presents the detection results.
Across all file types, FAV-based detection maintains near-perfect accuracy even when encryption is sparse and scattered.
In contrast, entropy- and $\chi^2$-based detectors rely on aggregate statistical deviation and therefore degrade significantly when encryption is applied to small, non-contiguous regions.
These results demonstrate that correctness with respect to file-format specifications provides a stronger and more reliable signal than byte-level statistical anomalies for detecting fine-grained partial-encryption attacks.

\begin{figure}[t]
    \centering
    \includegraphics[width=0.45\textwidth]{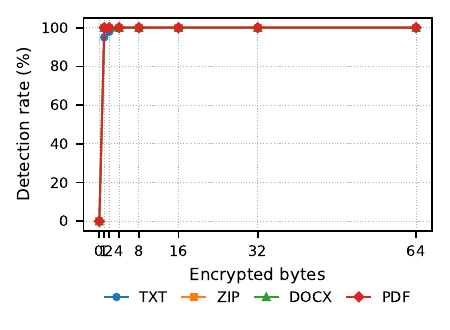}
    \vspace{-0.7cm}
    \caption{\textbf{Pushing FAV to the limit under fine-grained partial encryption (bytes $<64$).}
    Detection rate as the number of encrypted bytes per write is varied while the skip distance is fixed (skip=128).
    All formats remain perfectly detectable for encrypted bytes greater than 4\,B; TXT is the only case that exhibits slight degradation at 1--2\,B.
    }
    \label{fig:fav_limit_lt64}
    \vspace{-0.5cm}
\end{figure}

We further demonstrate that FAV-based detection is resilient to extremely fine-grained partial encryption by varying the encryption granularity below 64\,B under the same \emph{Skip-Step} attack pattern.
Figure~\ref{fig:fav_limit_lt64} shows that even in this extreme regime---down to single-byte encryption---FAV achieves 100\% detection for PDF, DOCX, and ZIP files.
Plain text (TXT) is the only corner case in which detection drops slightly, to 95.0\% at 1\,B and 97.86\% at 2\,B.
This behavior suggests that extremely small perturbations can occasionally evade UTF-8 validity checks due to statistical coincidence, where encrypted bytes happen to fall within valid UTF-8 encoding ranges.

\subsection{Detection of Real-World Attacks}

\begin{table}[t]
\centering
\caption{78 real-world ransomware samples are evaluated.}
\label{tab:realworld-ransomware-family}
\renewcommand{\arraystretch}{1.05}

\begin{tabular}{l c | l c}
\toprule
Family & Samples (\%) & Family & Samples (\%) \\
\midrule

Play           & 1 (1.28)  & LockFile       & 2 (2.56) \\
BlackBasta     & 1 (1.28)  & Cryptor        & 1 (1.28) \\
BlackMatter    & 2 (2.56)  & Chaos          & 1 (1.28) \\
TeslaCrypt     & 1 (1.28)  & GarrantDecrypt & 1 (1.28) \\
Hive           & 1 (1.28)  & Conti          & 1 (1.28) \\
Maze           & 1 (1.28)  & Mole           & 1 (1.28) \\
Akira          & 5 (6.41)  & Pandora        & 1 (1.28) \\
Medusa         & 3 (3.85)  & Vigorf         & 1 (1.28) \\
Rhysida        & 1 (1.28)  & Darkbit        & 1 (1.28) \\
AvosLocker     & 2 (2.56)  & Saturn         & 1 (1.28) \\
Virlock        & 1 (1.28)  & RagnarLocker   & 1 (1.28) \\
NightSky       & 1 (1.28)  & Nemty          & 1 (1.28) \\
Sodinokibi     & 2 (2.56)  & Ragnarok       & 1 (1.28) \\
Babuk          & 25 (32.05)& WastedLocker   & 1 (1.28) \\
Phobos         & 1 (1.28)  & Crysis         & 1 (1.28) \\
HelloKitty     & 1 (1.28)  & Yurei          & 1 (1.28) \\
VoidCrypt      & 1 (1.28)  & Lynx           & 1 (1.28) \\
Nitrogen       & 2 (2.56)  & Miscellaneous  & 10 (12.82) \\
\bottomrule
\end{tabular}
%\vspace{2pt}
\footnotesize
\end{table}

\scbf{Environment Setup} Our experiments for detecting real-world attacks were conducted on the same hardware platform described earlier. We used a dataset containing 140 text files, which were encrypted by each ransomware sample. Real-world ransomware samples were executed inside VirtualBox~7.2.4, and the system state was rolled back to a clean snapshot after each ransomware completed its encryption process.

\scbf{Attack Scenarios}
To obtain real-world ransomware samples, we collected 78 active evasive ransomware binaries from the MarauderMap dataset~\cite{MarauderMap2025} and MalwareBazaar~\cite{AbuseChBazaar2025}. For each sample, we assigned a ransomware family based on the label most frequently inferred by multiple detectors on VirusTotal~\cite{virustotal26}. Each sample was first validated in isolation before being executed inside the VM. Once verified, the ransomware was allowed to run until its encryption phase completed. After execution, we copied the encrypted files to the host over the network using the \texttt{scp} command, and the VM was rolled back to its initial clean state before running the next sample. FAV-based detection was then performed on the host using the copied encrypted files, and the results were analyzed using the detector output.

\scbf{Detection Results}
We evaluate how well \RansomSaver{} detects real-world ransomware attacks; the results are summarized in Table~\ref{tab:realworld-ransomware-family}. We map samples to 35 real-world ransomware families, while samples that do not map cleanly to a well-established ransomware family are labeled as \emph{Miscellaneous}. Across all evaluated ransomware samples, \RansomSaver{} correctly identifies all encrypted files with zero false positives and zero false negatives, achieving 100\% accuracy, precision, recall, and F1-score. These results demonstrate that \RansomSaver{} reliably detects real-world ransomware activity with perfect sensitivity and precision under the evaluated conditions.

\section{Discussion}
\label{sec:discuss}

Several widely used file formats deliberately permit regions whose syntax or
semantics are only weakly constrained and may legally contain arbitrary byte
sequences. Examples include non-stream gaps between objects, alignment padding,
incremental update sections, and auxiliary metadata in PDF files, as well as
prefix data, padding, or unclaimed gaps outside active entries in ZIP archives.
While this permissiveness supports flexible authoring and editing workflows, it
also creates opportunities for concealing small amounts of encrypted or
malicious data in a manner that remains syntactically valid and difficult to
distinguish from benign formatting artifacts.

This limitation is inherent to the base specifications of formats such as PDF~\cite{iso32000-1}
and ZIP~\cite{pkware-zip}, which do not require all bytes in a valid file to be semantically
accounted for or protected by an intrinsic integrity mechanism. As a result,
an attacker can embed ciphertext in these permissive regions without violating
format correctness. For instance, the PDF specification allows arbitrary binary
data in certain comment lines or unused gaps, while ZIP archives may contain
bytes not referenced by the central directory or covered by entry-level
integrity checks. When operating under the baseline specifications alone,
these regions remain residual attack surfaces.

In contrast, OOXML formats (\texttt{.docx}, \texttt{.xlsx}, \texttt{.pptx}),
defined by the Open Packaging Conventions (OPC)~\cite{iso29500-2}, impose stronger semantic
closure. All meaningful content must reside in explicitly typed package parts,
be referenced through well-defined relationships, and have a determinable
content type. As a result, OOXML packages do not admit the same class of
semantically unconstrained regions: bytes that fall outside the OPC structure
constitute structural violations rather than benign artifacts. Consequently,
the hiding surfaces present in permissive formats such as PDF and ZIP are
largely absent in well-formed OOXML documents.

To mitigate the risks posed by permissive regions in formats such as PDF and
ZIP, we impose additional restrictions beyond the base specifications in
security-sensitive settings. We capture this approach through a
\emph{clean-format contract} established between the detector and compliant
authoring or packaging tools. Under this contract, files are expected to avoid
unrestricted gaps or opaque regions unless protected by explicit integrity
mechanisms, such as cryptographic checksums, trusted manifests, or whitelists.
Any deviation from these stricter invariants—whether due to retained legacy
content, unconstrained padding, or unaccounted-for regions—is treated as a
contract violation and conservatively flagged as a potential integrity or
security issue. When such a contract cannot be assumed, \RansomSaver{}
acknowledges these permissive regions as inherent residual attack surfaces of
loosely specified formats.

% Attackers can also evade \RansomSaver{}'s uniformity-based detection in SAWA by encoding ciphertext into ASCII-armored representations such as Base64 or hexadecimal. These encodings restrict bytes to a small printable alphabet (e.g., \texttt{A--Z}, \texttt{a--z}, \texttt{0--9}, \texttt{+}/\texttt{/}), yielding lower per-byte entropy and producing data that appears similar to benign UTF-8 text. 
% To mitigate this evasion, regions containing Base64- or hex-like character sequences can be detected, decoded, and then re-evaluated using entropy, magic-byte, or format-aware validation to recover the underlying ciphertext if present. LLM-based semantic analysis may additionally be used to determine whether the decoded content is well-formed and contextually appropriate, helping to distinguish benign encoded data from covertly embedded ciphertext.
Attackers can also evade \RansomSaver{}'s uniformity-based detection in SAWA by
encoding ciphertext into ASCII-armored representations such as Base64 or
hexadecimal, which map arbitrary binary data into a restricted set of printable
ASCII characters. Because these encodings limit bytes to a small alphabet (e.g.,
\texttt{A--Z}, \texttt{a--z}, \texttt{0--9}, \texttt{+}/\texttt{/}), the resulting
data exhibits lower per-byte entropy and can resemble benign UTF-8 text rather
than uniformly random ciphertext.
To mitigate this evasion, regions containing Base64- or hex-like character
sequences can be detected, decoded, and re-evaluated using entropy, magic-byte,
or format-aware validation to recover the underlying ciphertext if present.
LLM-based semantic analysis may additionally be used to assess whether the
decoded content is well-formed and contextually appropriate, helping distinguish
benign encoded data from covertly embedded ciphertext.

Furthermore, modern ransomware increasingly engages in double extortion, threatening to leak victims' private data after exfiltrating it from compromised devices. This trend bypasses defenses that focus solely on detecting local encryption activity, including \RansomSaver{}. Effective mitigation requires restricting outbound data transfers through network egress filtering, monitoring for abnormal compression or upload behavior, and enforcing least-privilege access controls. \RansomSaver{} operates orthogonally to such mitigations, making future extensions toward exfiltration detection straightforward.

% Also, \RansomSaver{} assumes the availability of compliant applications
% that support scheduled or checkpoint-based rewriting of files to minimize false positives.
% While this may appear restrictive, both enterprise and consumer environments
% already rely on periodic update or maintenance windows (e.g., nightly backups,
% auto-save consolidation, or scheduled software updates).
% Integrating \RansomSaver{} into these existing routines therefore represents a
% practical extension of current practices rather than a deployment barrier.
\RansomSaver{} assumes the availability of applications that perform file
updates in a consolidated manner at well-defined points in time (e.g., save,
commit, or application-level checkpoint boundaries).
This assumption does not require new application semantics; rather, it aligns
with common deployment practices in both enterprise and consumer environments.
Many systems already rely on periodic consolidation or maintenance phases,
including nightly backups, auto-save flushing, and scheduled software updates.
\RansomSaver{} leverages these existing windows to perform validation, making
deployment a practical extension of current workflows rather than a barrier to
adoption.

\section{Conclusion}\label{sec:conclude}

% In this study, we presented \RansomSaver{}, a cloud-offloaded ransomware defense
% system that detects privilege-escalated and evasive attacks using snapshot-based,
% format-aware content validation.
% By grounding detection in file-format semantics rather than coarse statistical
% signals, \RansomSaver{} robustly identifies fine-grained partial-encryption
% attacks that intentionally evade entropy- and distribution-based detectors.
% Our evaluation shows that \RansomSaver{} achieves near-perfect detection accuracy
% for structured formats such as TXT, PDF, OOXML, and ZIP, even under extreme
% micro-granularity encryption, while operating with practical overhead and no
% specialized hardware.
% These results demonstrate that specification-level correctness provides a strong
% and scalable foundation for cloud-assisted ransomware defense.
In this paper, we presented \RansomSaver{}, a cloud-assisted ransomware defense
system that detects privilege-escalated and evasive attacks using snapshot-based,
format-aware content validation.
By grounding detection in specification-level correctness rather than
entropy- or distribution-based signals, \RansomSaver{} identifies fine-grained
partial-encryption attacks that evade statistical detectors.
Our evaluation demonstrates near-perfect detection accuracy across structured
formats including TXT, PDF, OOXML, and ZIP, even under micro-granularity
encryption, while incurring practical overhead.
These results show that format-aware, specification-driven validation provides a
scalable foundation for defending against modern ransomware attacks.

\begin{acks}
%This paper was edited for grammar using ChatGPT.
\end{acks}

% \scbf{Generative AI Usage}
% We used ChatGPT as an assistive tool during this research. Specifically, ChatGPT was used for (i) minor editorial improvements, including grammar correction, wording refinement, and stylistic polishing of the manuscript, and (ii) generating initial code drafts based on detailed, human-written design specifications. All AI-assisted code was subsequently reviewed, revised, and validated by the authors through manual inspection, unit testing, integration testing, and end-to-end experimental evaluation. All experimental results, data, and conclusions reported in this paper were produced, verified, and interpreted by the authors. The authors take full responsibility for the correctness, originality, and integrity of the work.

\bibliographystyle{ACM-Reference-Format}
%\bibliography{sample-base}
\bibliography{bk}

\appendix

\end{document}